\documentclass{article}
\usepackage{fullpage}
\usepackage{amsmath}
\usepackage{amsfonts}
\usepackage{amssymb}
\usepackage{graphicx}

\newcommand{\im}{{\mbox{Im}\,}}

\newcommand{\ima}{\hbox{Im}\,}
\newcommand{\rea}{\hbox{Re}\,}

\newcommand{\sa}{s_2}

\newcommand{\mo}{O}
\newcommand{\disc}{\hbox{Disc}\,}
\newcommand{\NP}[1]{Nucl.\ Phys.\ {#1}}

\newcommand{\PR}[1]{Phys.\ Rev.\ {#1}}

\begin{document}
\title{The Inverse Amplitude Method and Adler Zeros}

\author{A. G\'omez Nicola, J.R. Pel\'aez and G. R\'ios \\
\emph{Departamento de F\'{\i}sica Te\'orica II.}\\
\emph{Universidad Complutense. 28040 Madrid. Spain}
}
\date{}
\maketitle

\begin{abstract}
The Inverse Amplitude Method is a powerful unitarization technique
to enlarge the energy applicability region of Effective Lagrangians.
It has been widely used to describe resonances from Chiral
Perturbation Theory as well as for the Strongly Interacting Symmetry
Breaking Sector. In this work we show how it can be slightly
modified to account also for the sub-threshold region, incorporating
correctly the Adler zeros required by chiral symmetry and
eliminating  spurious poles. These improvements  produce negligible
effects on the physical region.
\end{abstract}

\section{Introduction}
Effective Field Theories  provide a systematic and model independent
approach to systems whose symmetries and low energy degrees of
freedom are known  but whose description in terms of an underlying
fundamental Quantum Field Theory is out of reach. The two cases of
interest for this work are, on the one hand, Chiral Perturbation
Theory (ChPT) \cite{ChPT} that describes effectively the low energy
dynamics of hadrons, unaccessible to perturbative QCD calculations
in terms of quarks and gluons and,   on the other hand,  the
effective description of the Strongly Interacting Electroweak
Symmetry Breaking Sector (SISBS) \cite{Appelquist:1980vg}, whose
underlying fundamental theory remains unknown.

Both cases have in common the existence of an spontaneous symmetry
breaking of a global chiral $SU(N)_L\times SU(N)_R$ group down to
an $SU(N)_{L+R}$ group. The Goldstone Theorem implies the presence
of $N^2-1$ massless Goldstone Bosons (GB) in the particle
spectrum. These GB thus become the relevant degrees of freedom of
the system below a chiral scale $\Lambda_\chi$, where an Effective
Chiral Lagrangian can be built  in terms of just those GB as the
most general derivative expansion consistent with the known
symmetries. Note that there is
 a well defined power counting so that divergences generated in
loop diagrams that use vertices up to a given order can be
renormalized into the coefficients of higher order terms. In this
sense these Effective Theories are renormalizable order by order.

In addition, in both examples, there are small explicit symmetry
breaking terms. For QCD the relatively small masses of the
lightest three quarks provide a mass $M<<\Lambda_\chi$ to the GB,
identified with the pions, kaons and etas, so that the effective
approach becomes in practice a derivative and mass expansion. For
the SISBS, there is a local $SU(2)_L\times U(1)$ symmetry whose
gauge bosons couple to the "`would-be GB"' that, in a unitary
gauge, disappear from the spectrum, giving rise to gauge boson
longitudinal components that thus acquire a mass $M_V$. In this
way the $SU(2)_L\times U(1)$ gauge symmetry is spontaneously, but
not explicitly, broken to the electromagnetic group $U(1)_{EM}$.
In this case, in addition to the derivative expansion, one expands
also in terms of electroweak coupling constants $g$ and $g'$. The
so-called Equivalence Theorem (ET) \cite{ET} states that at high
energies (in $R_\xi$ gauges, and to leading order  in momenta over
$M_V$ and $g$ and $g'$) amplitudes involving longitudinal gauge
bosons can be calculated as if they were GB, which, being
pseudoscalars, are much easier to handle. Although this is a high
energy limit, there is a generalization to the Effective
Lagrangian formalism \cite{ETnosotros}, that, for practical
purposes, allows us to identify, up to the difference in scales,
the formalisms of $SU(2)$ ChPT and the SISBS and therefore, from
now on we will be referring to ChPT, but keeping in mind that our
results have a straightforward translation to the SISBS.

Both cases above are examples of strongly interacting systems
whose most salient feature is the saturation of unitarity and the
associated resonant states, which lie beyond the reach of
perturbative energy expansions. Thus, it may seem that the use of
effective Lagrangians is limited to energies below those
resonances, whose effects are encoded in the values of higher
order effective coupling constants. However, since unitarity fixes
the imaginary part of $\it inverse$ partial waves in the elastic
region,  the effective Lagrangian approach is also useful in the
resonant region, for instance, used inside a dispersion relation,
in order to obtain the rest of the amplitude. These techniques are
known as unitarization methods, and reproduce simultaneously the
low energy expansion and the lightest resonances without including
them explicitly in the Lagrangian. The great advantage is that
such resonances and their properties are generated without
prejudices about their nature or their existence. Also, since the
Lagrangian symmetries and some features of the effective constants
can be directly related to the underlying theory, like QCD, one
can study the properties of these states based on more fundamental
grounds. One of the most extensively used unitarization techniques
is the Inverse Amplitude Method (IAM)
\cite{Truong:1988zp,Dobado:1992ha,Dobado:1996ps,Guerrero:1998ei},
that uses the fully renormalized
 effective chiral expansion, without any further
approximation and without introducing any other spurious parameter,
but just the effective constants up to a given order. Within hadronic physics,
it generates the well known vector resonances and the more controversial scalars
using parameters consistent 
with one loop ChPT, allowing to establish their different
nature in terms of their dependence on the number of colors. It has also been possible to
extend it up to two loops \cite{Dobado:1996ps,Nieves:2001de},
the finite temperature formalism \cite{Dobado:2002xf,FernandezFraile:2007fv}
and to the pion-nucleon sector \cite{Pelaez:1999ey}.
Within the SISBS \cite{Dobado:1989gr}, it provides the prediction of the general
resonance spectrum and how well it could be detected at LHC.

However, it is known \cite{Dobado:1996ps,Boglione:1996uz} that the
IAM fails to reproduce correctly the Adler zeros that appear in
the subthreshold region of some partial waves as a consequence of
chiral symmetry \cite{Adler:1965ga}. Furthermore, it generates spurious poles, or
``ghosts'', thus questioning its reliability in that region, and
also casting some doubts about the robustness of the results in
the physical region if such structures were properly accounted
for.

The aim of this paper is to show that a very simple modification of the
IAM can take into account correctly those zeros and ghosts. This modification corresponds to 
terms that had been neglected in the original dispersive derivation of the IAM since they contribute to higher orders in the chiral expansion.
Actually, we will check explicitly that such procedure
 is justified in the physical region, where these modifications yield negligible contributions, thus showing the robustness of the standard IAM.
However, apart from improving the IAM consistency, these terms are essential
in the sub-threshold region which is relevant to
study the effect of chiral symmetry restoration on resonances
\cite{FernandezFraile:2007fv}, or their
dependence on quark masses \cite{inprep}.

In the next section we will thus revisit the standard IAM derivation
from dispersive theory, where Adler zeros are neglected, paying
special attention to the role of those zeros in the subthreshold
region.  In section \ref{sec:naive} we will present a naive way of
extending the IAM amplitude, without using dispersion relations,
that solves the caveats in that region. Section \ref{sec:dispequal}
will show a dispersive derivation  of a more general modified
amplitude, for the case of equal masses (e.g. $\pi\pi$ scattering).
The case of unequal masses, like in $\pi K$ scattering, deserves a
separate discussion, for the reasons explained in section
\ref{sec:uneq}. Finally, in section \ref{sec:num} we will present
some numerical results for the modified amplitudes.

\section{The  Inverse Amplitude Method}

\subsection{Dispersive derivation}

\begin{figure}[h]
  \centering
  \includegraphics[scale=.4]{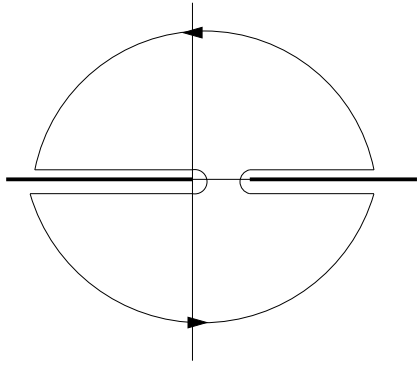}
  \caption{ \label{contour}
 Pion-pion scattering partial waves analytic structure and
integration
  contour.}
\end{figure}

The {\it one-channel} Inverse Amplitude Method (IAM)
\cite{Truong:1988zp,Dobado:1992ha,Dobado:1996ps,Nieves:2001de} can
be obtained by using ChPT up to a given order inside a dispersion
relation. To simplify the discussion, let us first consider
pion-pion scattering partial wave amplitudes of definite isospin $I$
and angular momentum $J$, although for brevity we will simply call
them $t$, whose analytic structure in the $s$-plane is shown in
Fig.~\ref{contour}. The physical right hand cut comes from unitarity
and starts at threshold $s_{th}$, while the left hand cut comes from
the $t$,$u$ channels.

The inverse of $t(s)$ has the same analytic structure, except for
the possible presence of poles corresponding to zeros of $t(s)$. In
particular, chiral symmetry requires the existence of the so-called
Adler zeros below threshold in scalar partial waves
\cite{ChPT,Adler:1965ga}, the case of interest in this work, which
we denote by $s_A$.  Hence it is then possible to write the
following dispersion relation for the inverse amplitude:
\begin{eqnarray}
    \label{1/tdisp}
  \frac1{t(s)}=\frac1{t(z_0)}+
    \frac{s-z_0}{\pi}\int_{s_{th}}^{\infty}\,dz\,
    \frac{\im\,1/t(z)}{(z-s)(z-z_0)}+
LC(1/t)+PC(1/t).
\end{eqnarray}
here and in the following, we will simply write $z$ instead of $z+i
\epsilon$ with $\epsilon>0$ for the imaginary parts inside the cut
integrals. Note that we have explicitly written the integral over
the right hand cut (or physical cut, extending from threshold,
$s_{th}$ to infinity) but we have shortened by $LC$ the equivalent
expression for the left cut (from $-\infty$ to 0) and the pole
contribution. We could do similarly with other cuts, if present, as
in the $\pi K$ case. In addition we have made one subtraction to
ensure convergence, at a point $z_0\neq s_A,s_2$.

We now recall that unitarity, for physical values of $s$ in the
elastic region implies:
\begin{equation}
  \ima t(s) =\sigma \vert t(s)\vert^2 \quad\Rightarrow \quad\ima \frac{1}{t(s)}=-\sigma(s),
\label{unit}
\end{equation}
where $\sigma(s)=2p_{CM}/\sqrt{s}$. Let us remark that since ${\rm
Im} 1/t=-\sigma$
 we know exactly the integrand
over the elastic cut.

In contrast, ChPT amplitudes are obtained as a series expansion
$t(s) =t_2(s)+t_4(s)+...$
where $t_2(s)=O(p^2)$, $t_4(s)=O(p^4)$ and $p$ stands for the
pion mass or momentum. Therefore elastic
unitarity is not satisfied exactly, but only order by order as follows:
 \begin{equation}
 \ima t_2(s)=0,\qquad \ima t_4(s)=\sigma(s)\vert t_2(s)\vert^2, \quad ...
 \label{pertunit}
 \end{equation}
Let us also note that $t_2(s)$ is a pure polynomial and has no cuts,
and we can thus write a trivial dispersion relation for $1/t_2(s)$  that reads
\begin{equation}
  \label{1/t2disp}
  \frac1{t_2(s)}=\frac1{t_2(z_0)}+PC(1/t_2),
\end{equation}
where now the pole contribution is due to $s_2$, the Adler zero of
$t_2$.  In addition, except for the poles, the function $t_4(s)/
t_2^2(s)$ has the same analytic cut structure of $1/t$ and, using 
Eqs.\eqref{unit} and \eqref{pertunit}, over the
physical cut we find
\begin{equation}
\ima \frac{t_4(s)}{t^2_2(s)}=\sigma(s)=-\ima \frac{1}{t(s)}.\label{eq:unitsigma}
\end{equation}

 We can therefore write
another dispersion relation similar to that of $1/t(s)$, but for $t_4(s)/ t_2^2(s)$,
\begin{equation}
  \label{t4/t2disp}
  \frac{t_4(s)}{t_2(s)^2}=\frac{t_4(z_0)}{t_2(z_0)^2}+
  \frac{s-z_0}{\pi}\int_{s_{th}}^{\infty}\,dz\,\frac{\im t_4(z)/t_2(z)^2}
  {(z-s)(z-z_0)}+LC(t_4/t_2^2)+PC(t_4/t_2^2),
\end{equation}
where the pole contribution, once again, is due to the Adler zero of
$t_2$.

We are now going to relate the dispersion relation for $1/t(s)$ with that for
$t_4(s)/ t_2^2(s)$. As we already commented $\im 1/t(s)=-\im t_4/t_2^2(s)$ on the right
cut, and therefore the integrals over the physical cuts for $1/t(s)$
and $t_4/t_2^2(s)$ {\it are exactly opposite to each other}. 
In addition, using ChPT we find that $LC(1/t)\simeq
-LC(t_4/t_2^2)$, which is a well justified approximation, since, due to the subtraction, 
the integrand of $LC$ is weighted at low energies, precisely where ChPT applies.
Finally, we have to evaluate the subtraction constant in Eq.\eqref{1/tdisp}, and this can only be done as long as $z_0$ is in the low energy region, where it is perfectly
justified to use ChPT to find $1/t(z_0)\simeq
1/t_2(z_0)-t_4(z_0)/t_2(z_0)^2$. However, note that this
expansion is a very bad approximation for $z_0$ near $s_2$ or $s_A$, where $t_2$ and $t$
vanish.  Therefore, we only know how to relate those dispersion relations 
for subtraction points $z_0$ in the low energy region, but sufficiently far from the Adler zeros. In section 6  we will check
that the results have very little sensitivity to the choice of $z_0$ as long as it lies
in this region. When this is the case, using
Eqs.\eqref{1/tdisp}, \eqref{1/t2disp} and \eqref{t4/t2disp}
 we can write $1/t$ as
\begin{equation}
  \label{IAMpc}
  \frac1{t(s)}\simeq\frac1{t_2(s)}-\frac{t_4(s)}{t_2(s)^2}
-PC(1/t_2)+PC(t_4/t_2^2)+PC(1/t)
\end{equation}

The standard dispersive derivation of the IAM
\cite{Dobado:1992ha,Dobado:1996ps}
simply neglected the sum of  pole contributions
to arrive at
\begin{equation}
t^{IAM}(s)\simeq
\frac{t^2_2(s)}{
t_2(s)-t_4 (s)},
\label{IAM}
\end{equation}
thus providing an elastic amplitude that satisfies unitarity and has
the correct low energy expansion of ChPT up to the order we have
used. When such amplitude is chirally expanded to  $O(p^4)$, this
implies \cite{Dobado:1996ps} that the total pole contribution to $t(s)$, even without its
explicit calculation, has to be $O(p^6)$. In sections
\ref{sec:naive},\ref{sec:dispequal},\ref{sec:uneq} we will calculate
it explicitly to arrive to the modified IAM.

\subsection{IAM properties and its naive derivation}

Incidentally, we can recast Eq.\eqref{unit} as:
\begin{equation}
  \label{eq:otra}
t(s)=\frac{1}{\rea t^{-1}(s) - i\, \sigma(s)},
\end{equation}
and thus  it seems that the IAM can also be derived in a much simpler way by replacing
$\rea t^{-1}$ by its $O(p^4)$ ChPT expansion $\rea t^{-1}=(t_2- {\rm
Re}\,t_4)/t_2^2$. This is the way unitarization methods are usually
presented, although it makes no use of the strong analytic constraints of amplitudes, 
which are indeed absent in Eq.\eqref{eq:otra}. Furthermore,
 immediately the criticism is raised that the ChPT
expansion cannot be used at high energies. 

However, note that the dispersive
one-channel IAM derivation in the previous section imposes analyticity
in the form of two dispersive
integrals and makes use of the ChPT expansion (up to one-loop in this case) for
the subtraction constants and the left cut. 
The use of ChPT for the
subtraction constants is well justified, since it is used at $s=z_0$
in the low-energy region. Since the integral extends to infinity,
ChPT may seem a worse approximation for the left cut or possible
inelastic cuts, however, the subtraction suppresses the high energy
region that contributes little, as explained above. Furthermore,
when the IAM is used for physical values of $s$ above the physical
threshold, the left cut is damped by an additional $1/(z-s)$ factor,
and not only the high energy part, but its whole contribution is
rather small. There are no model dependent assumptions, but just
{\it approximations to a given order}, and therefore the approach provides
an elastic amplitude that satisfies unitarity and has the correct
ChPT expansion up to that given order. It also is straightforward to
extend it to higher orders \cite{Dobado:1996ps,Nieves:2001de}.

Moreover
 Eq.\eqref{unit} is only valid on the physical cut,
whereas the dispersive
derivation allows
us to consider the amplitude in the complex plane, and,
in particular, look for poles of the
associated resonances. Actually,
already ten years ago \cite{Dobado:1996ps} the poles for the
$\rho(770)$, $K^*(892)$ and most interestingly,
the controversial $\sigma$
(also called $f_0(600)$), where generated
without any model dependent assumptions.

Obviously, and contrary to a wide belief, the IAM {\it contains a
left cut} and {\it respects crossing symmetry} up to, of course, the
order in the ChPT expansion that has been used. The confusion may
come from the fact that the IAM has also been applied in a coupled
channel formalism, for which {\it there is still no dispersive
derivation} and has been frequently used by approximating further
the amplitudes neglecting tadpole and left cut terms
\cite{Oller:1997ng}. But, strictly speaking that would not be the
full IAM, which definitely has a left cut.

\subsection{IAM caveats: Adler zeros and ghosts}

To end this section, we recall that in the dispersive derivation 
the sum of pole terms (PC) in Eq.(\ref{IAMpc}) is neglected, since it yields a
higher order contribution \cite{Dobado:1996ps}. However, this leads
to a couple of problems related to the presence of the Adler zeros
below threshold in the scalar waves. 
Let us first note that despite we have used in the IAM 
the ChPT expansion up to next to leading order (NLO), due to the $t_2(s)^2$ factor
in the numerator, it vanishes at $s_2$, which is only the leading order (LO)
chiral approximation 
to the exact Adler zero $s_A$. 
In addition it is a
double zero instead of a single zero. This, as we will see, leads to
the appearance of a spurious pole on the real axis close to the
Adler zero. As a consequence, the predictions of the standard IAM below
threshold and in particular, around the Adler zero position, are not
reliable.

In fact, note that since the interval $s\in(0,s_{th})$ lies within
the  convergence region of the chiral expansion and $t_2$ changes
sign at $s=\sa$,  $t_2-t_4$ turns out to be, for the cases of
interest here, a continuous, monotonically increasing (or
decreasing) function in $(0,s_{th})$ that changes sign from $s=0$ to
$s=s_{th}$. Therefore, there is one single point $\tilde s$ where
the denominator of (\ref{IAM}) vanishes, which, as long as
$t_4(s_2)\neq 0$, produces the spurious pole below threshold
discussed above, i.e., a not-observed $\pi\pi$ bound  state.

For instance, let us consider $I=J=0$ $\pi\pi$ scattering (for the values of the
low-energy constants quoted in section \ref{sec:num}. This is an
attractive channel, so that $t_2-t_4$ is positive at threshold.
Since that function is negative at $s=0$ and at $s=\sa$,  $\tilde
s>\sa$ in that case, as showed in the upper right panel of Fig.\ref{fig:comparison}, 
we find $\tilde s\simeq (110 \ \mbox{MeV})^2$  and $\sa\simeq
(99 \ \mbox{MeV})^2$. In the $I=2,J=0$ channel, which is repulsive,
$t_2-t_4$ is positive at $s=0$ and negative at $s=\sa$
($t_4(\sa)>0$) so that $\tilde s<\sa$. In that channel, $\tilde
s\simeq (196 \ \mbox{MeV})^2$ and $\sa\simeq (198 \
\mbox{MeV})^2$ as seen also in Fig.\ref{fig:comparison}.

Let us also point out that, together with the first Riemann sheet
spurious pole just discussed, the IAM has a companion
pole in the second Riemann sheet below  threshold. For instance,
the second-sheet IAM for $\pi\pi$ scattering reads:

\begin{equation}
  t(s)^{IAM,II {\rm sheet}}= \frac{t_2(s)^2}{t_2(s)-t_4(s)-2\tilde\sigma(s) t_2^2(s)}
  \label{iamsecrs}
\end{equation}
where
$\tilde\sigma(s)=i\sigma(s-i0^+)=\sqrt{4m_\pi^2/s-1}$  for
$0<s<4m_\pi^2$. Thus, if we are dealing with an attractive
channel, like the $00$ one, the denominator of (\ref{iamsecrs}) is
positive near threshold (dominated by $t_2> 0$) and diverges to
minus infinity as $s\rightarrow 0^+$, so that it must have at
least one zero, which again generates a pole if $t_4(\sa)\neq 0$.
Since
$-\tilde\sigma t_2^2$  in Eq.(\ref{iamsecrs})
is also an increasing
function, there will be only one such pole.

In conclusion, the existence of the perturbative Adler zeros and the
fact that the IAM amplitude does not reproduce them well leads to
the presence of spurious poles. Similar conclusion had been noticed
in \cite{bugg03}. In the next sections, we present, first, a very
simple construction of a modified IAM, along the lines of the
previously discussed naive derivation of the standard IAM obtained
without using dispersion relations, that solves these problems and
next we show two dispersive derivations of the modified IAM. One of
them is subtracted at arbitrary $z_0$ (within the range of validity
of our approximations) and the other one at the Adler zero. We will
show how the modified IAM obtained naively corresponds to a
particular limit of the first dispersive approach and comes out
directly in the second. Finally, we will also show that the
differences between the modified IAM formulae are negligible
numerically, and that, while fixing the above mentioned problems, the
modified IAM does not yield any significant modification over the
standard IAM in the physical region and to the resonance poles.
Therefore, the results obtained so far in the literature with the
IAM remain valid.

\section{Modified IAM: Naive derivation}
\label{sec:naive}

First of all we will set some notation: as before, we denote by
$s_A$ the Adler zero of the ``complete'' partial wave, namely
$t(s_A)=0$. In addition, we will also use the approximations to the
Adler zero at LO, $s_2$, and NLO,
$s_2+s_4$. Thus, $t_2(s_2)=0$ and $t_2(s_2+s_4)+t_4(s_2+s_4)=0$.

In this section we present a naive, and intuitive, derivation leading to a partial
wave definition which does not have the Adler zero related
problems discussed above. The derivation follows closely
\cite{FernandezFraile:2007fv}, where this problem was addressed in
the context of real pion scattering poles arising by medium
effects such as temperature and density. In that paper, there was
not a formal proof based on dispersion relations, as the one we
will present here later, and it was limited to pion-pion scattering.

>From the discussion in the previous sections, it is clear that if we
modify the inverse amplitude as $1/t^{IAM}(s)\rightarrow
1/t^{IAM}(s)+A(s)/t_2^2$ with $A(s)$ an analytic function at least off the
real axis, real for real $s$, the unitarity and analytic properties
of the amplitude remain unaltered. The modified IAM, (from now on called mIAM), reads
then:

\begin{equation}
t^{mIAM}(s)= \frac{t^2_2(s)}{ t_2(s)-t_4 (s)+A^{mIAM}(s)}, \label{IAMext}
\end{equation}

Consider now the case of $\pi\pi$ scattering, where we have
simply $t_2(s)=t'_2(s_2)(s-s_2)$ with $t'_2(s_2)$  constant. Then
the Laurent expansion around $s=s_2$ of the standard IAM
reads:

\begin{equation}
  \frac{1}{t^{IAM}(s)}=
  -\frac{t_4(s_2)}{{t'_2(s_2)}^2 (s-s_2)^2}+\frac{t'_2(s_2)-
    t_4'(s_2)}{{ t'_2(s_2)}^2(s-s_2)}+\mo (s-s_2)^0.
  \label{Laurent1/t}
\end{equation}
The idea is that if we want the amplitude  to
 have \emph{only} an Adler
zero of order one at $s=s_A$, we must subtract from $1/t^{IAM}$
the above double and single pole contributions at $s=s_2$ and add
a single pole at  $s_A$, i.e,
\begin{equation}
\label{IAM+cosas}
  \frac1{t^{mIAM}(s)}=\frac1{t^{IAM}(s)}+
  \frac{t_4(s_2)}{{t'_2(s_2)}^2 (s-s_2)^2}-
  \frac{t'_2(s_2)-t_4'(s_2)}{{ t'_2(s_2)}^2(s-s_2)}+
  \frac{c}{s-s_A},
\end{equation}
where $c$ is a so far undetermined constant that, as we will show
now, can be fixed by demanding that the mIAM formula matches
the perturbative ChPT series to fourth order, namely $A=\mo(p^{6})$.

In practice, it is simpler to keep track of the different chiral powers by counting the  powers of $f^{-2}$, where $f$ is the
pion decay constant. Thus, since $t'_2(s_2)=\mo(f^{-2})$, $t_4=\mo(f^{-4})$,
 and $s_4=\mo(f^{-2})$,
expanding the expression $t_2(s_2+s_4)+t_4(s_2+s_4)=0$ around
$s_2$ we find:
\begin{equation}
  s_4=-t_4(s_2)/t'_2(s_2)+\mo(f^{-4}).
\end{equation}
Using this in Eq.(\ref{IAM+cosas})
with $s_A=s_2+s_4+\mo(f^{-4})$, and requiring that
Eq.\eqref{IAMext} matches the chiral expansion at low energies,
we find the first two orders of the chiral
expansion for $c$:
\begin{equation}
  c=\frac1{t'_2(s_2)}-\frac{t_4'(s_2)}{t'_2(s_2)^2}+\mo(f^{-2}),
\label{cchoice}\end{equation}
which leads to
\begin{equation}
  \label{Aangelpipi}
  \begin{aligned}
    A^{mIAM}(s)=t_4(s_2)-\frac{(s_2-s_A)(s-s_2)}{s-s_A}\,
    \left[t'_2(s_2)-t'_4(s_2)\right].
  \end{aligned}
\end{equation}

Therefore, the mIAM in Eq.(\ref{IAMext}) with $A^{mIAM}(s)$ in
Eq.(\ref{Aangelpipi}) and $s_A$ approximated by its chiral expansion
given above,  matches the chiral expansion of the amplitude up to
fourth order and has the Adler zero at the same position and with
the same order as the perturbative amplitude. Furthermore, we have
solved in turn the spurious pole problem. In fact, since
$A^{mIAM}(s_2)=t_4(s_2)$, the denominator of Eq.(\ref{IAMext}) vanishes at
$s=s_2$. But, for $s\neq s_2$ we have shown that $A^{mIAM}(s)=\mo(f^{-6})$ and
therefore our previous argument about the monotonous behavior of
the denominator still holds, so that the denominator vanishes {\em
only} at $s=s_2$, where there is no pole contribution. The same
holds for the spurious second-sheet poles. In Fig.~\ref{fig:comparison} 
we show the mIAM amplitude in the $I=J=0$
channel, which we observe that is not singular below threshold and
remains close to the standard IAM result away from the Adler zero
region. The same situation takes place in the $I=2,J=0$ channel.
Finally, we  have checked that, as expected from
our previous arguments, the $f_0(600)$ or $\sigma$ pole
remains at the same place either using the second Riemann sheet extensions
of the mIAM or the IAM (see Table \ref{poles}).

In the next sections, we will check how a modified
IAM can also be obtained
by considering explicitly the
pole contributions in the dispersive derivation,
thus ensuring the correct analytic properties of the amplitude.
We will also show that the modified formula
obtained with the naive derivation in this section
can also be obtained as a particular case.

\section{Modified IAM: dispersive derivation
for equal masses.} \label{sec:dispequal}

\subsection{Pole Contribution to the standard derivation}

The derivation of the modified IAM from dispersion theory follows
that in section 2 up to Eq.\eqref{IAMpc}, but keeping the pole
contributions, which, by evaluating the corresponding residues,
read:
\begin{eqnarray}
  PC(1/t_2)&=&
  \frac1{t'_2(s_2)}
  \left(
    \frac1{s-s_2}-\frac1{z_0-s_2}
  \right)
  \label{eq:PC1/t2}\\
  PC(t_4/t_2^2)&=&
  \frac{t_4(s_2)}{t'_2(s_2)^2}
  \left(
    \frac1{(s-s_2)^2}-\frac1{(z_0-s_2)^2}
  \right)+
    \frac{t'_4(s_2)}{t'_2(s_2)^2}
    \left(
      \frac1{s-s_2}-\frac1{z_0-s_2}
      \right)
  \label{eq:PCt4/t2}\\
  PC(1/t)&=&
  \frac1{t'(s_A)}
  \left(
    \frac1{s-s_A}-\frac1{z_0-s_A}
  \right)
  \label{res1/t}
\end{eqnarray}
where we have assumed a {\it single} zero in $t(s_A)$, as the
presence of the $1/(s-s_A)$ factor in $PC(1/t)$
shows. As we have discussed above, and as it is detailed in the
Appendix, by expanding chirally $-PC(1/t_2)+PC(t_4/t_2^2)+PC(1/t)$,
the poles contribute to $t(s)$ at higher order, and that is why they
were customarily neglected. However, these pole contributions
contain the terms needed to have the Adler zero in the correct
position. In addition, by taking the limit $s\rightarrow s_2$,
$(-PC(1/t_2)+PC(t_4/t_2^2))$ tends to the term needed to cancel the
double zero of the IAM in $s_2$, see Eq.\eqref{Laurent1/t}, and  the
spurious pole will also disappear.

In summary, the modified IAM obtained from dispersive relations subtracted at $z_0$,
that we will denote by $z_0IAM$,  can be
written again as:
\begin{equation}
t^{z_0IAM}(s)\simeq
\frac{t^2_2(s)}{
t_2(s)-t_4 (s)+A^{z_0IAM}(s)},
\label{IAMmod}
\end{equation}
where now
\begin{equation}
  \label{Az0pipi}
  \begin{aligned}
    A^{z_0IAM}(s)=A^{mIAM}(s)-\frac{t_2(s)^2}{t_2(z_0)^2}A^{mIAM}(z_0)
  \end{aligned}
\end{equation}
Of course, the position of the Adler zero for the total amplitude is
not known, but since we have been working with ChPT to one loop, we
can impose the Adler zero to be located in its one-loop position,
namely, in the above formulae, we have to replace $s_A\to s_2+s_4$
which is obtained from the equation $t_2(s_2+s_4)+t_4(s_2+s_4)=0$ as
explained in Section \ref{sec:naive}. Thus, to obtain
(\ref{Az0pipi}) we have made use of
$t_2(s)=t'_2(s_2)(s-s_2)$ and chirally expanded 
$1/t'(s_A)\simeq 1/t'_2(s_2)-t'_4(s_2)/t'_2(s_2)^2$, 
which is perfectly justified near $s_A$. Note that we can use the chiral expansions around
$s_A$ and $s_2$ because we no longer expand the inverse of 
the amplitudes but that of their derivatives,
as they appear in the residues of the pole contributions.

Note that, once again, the factor $A^{mIAM}(s)$ 
that appeared in the previous section
is present, but now there is an additional and very similar piece that carries 
a $z_0$ dependence, which occurs due to our truncation of the ChPT series when
approximating the subtraction constants and pole contributions. 
This additional term in $A^{z_0IAM}$ is also $O(f^{-6})$ but, as
we will see below, as long as $z_0$ lies within
the range where our approximations remain valid, it is numerically small 
not only in the physical region but also below threshold. Therefore, this
term can be dropped without spoiling the right chiral behavior of
the amplitude in the subthreshold region, obtaining again the mIAM in section \ref{sec:naive}, which, for the moment, we have justified only numerically.

Furthermore,  it is tempting to take the  $z_0\to\infty$ limit
in Eq.~\eqref{Az0pipi} and recover the mIAM of the previous section by
noting that the second term in Eq.\eqref{Az0pipi} vanishes in that limit.
However, this is just a formal justification of our
naive derivation
since, actually, we required for the subtraction point 
$z_0$ to remain in the low energy 
applicability region of ChPT. 
 Nevertheless, we will see in the next section that there
is an alternative dispersive
derivation of the mIAM, which is somewhat different from
the standard dispersive derivation since it requires subtractions at the
Adler zero of each function.

\subsection{Subtraction at the Adler zeros}
\label{subsAdlerpipi}
The way to derive Eq.\eqref{Aangelpipi} from dispersive theory is to make the
subtractions at the place where we already have a pole. Note,
however, that $1/t$ has its pole in the Adler zero at $s_A$
whereas $t_4/t_2^2$ has its pole at $s_2$, so that we have to write
\begin{equation}
  \label{1/tdispA}
  \frac1{t(s)}=-\frac{s-s_A}{\pi}\int_{RC}dz\,
  \frac{\sigma(z)}{(z-s_A)(z-s)}+
  LC(1/t)+PC(1/t),
\end{equation}
\begin{equation}
  \label{t4/t2dispA}
  \frac{t_4(s)}{t_2(s)^2}=\frac{s-s_2}{\pi}\int_{RC}dz\,
  \frac{\sigma(z)}{(z-s_2)(z-s)}+
  LC(t_4/t_2^2)+PC(t_4/t_2^2),
\end{equation}
where, for brevity, we have already used the elastic unitarity
condition Eq.\eqref{eq:unitsigma}. 
As usual, we will approximate $LC(1/t)\simeq -LC(t_4/t_2^2)$,
since that is the result of the chiral
expansion at low energies where
the integral is weighted. In the above relations, the pole
contributions $PC(1/t)$ and $PC(t_4/t_2^2)$ come now from a double
and triple pole respectively, and read
\begin{equation}
  \label{PCsA}
  PC(1/t)=\frac1{t'(s_A)(s-s_A)}-\frac{t''(s_A)}{2t'(s_A)^2},
\end{equation}
\begin{equation}
  \label{PCs2}
  PC(t_4/t_2^2)=
  \frac{t_4(s_2)}{t'_2(s_2)^2(s-s_2)^2}+
  \frac{t'_4(s_2)}{t'_2(s_2)^2(s-s_2)}+
  \frac{t''_4(s_2)}{2t'_2(s_2)^2}.
\end{equation}
In addition, it is important to remark that these
pole contributions diverge either on
$s_2$ or $s_A$, so that, for $s\simeq s_2$
or $s\simeq s_A$ they are, by far, the dominant contributions, since at the same time
the right and left cut terms tend to zero,
and therefore become negligible. 

Outside that region, the other terms become relevant and 
the difference with our previous derivations is that
 now we also approximate the $1/t$ integral
term over the right cut
by using $(s-s_A)/(z-s_A)\simeq (s-s_2)/(z-s_2)$, which is its
LO chiral expansion. This is a remarkably
good approximation for the dispersion relation as long as
$z$ is sufficiently far from $s_2$ and $s_A$,
which is indeed the case right cut integral.
Of course, the $1/t$ right cut term should vanish
at $s_A$ and now it does not, but, as we have just commented,
the pole contribution diverges
precisely at $s_A$ and thus is largely dominant
over the integral, which therefore can be completely neglected.

Once again, we simply add Eqs.\eqref{1/tdispA} and
\eqref{t4/t2dispA} to obtain the mIAM equation:
\begin{equation}
  \label{preIAM-A}
  \frac1{t^{mIAM}(s)}=-\frac{t_4(s)}{t_2(s)^2}+
  \frac{t_4(s_2)}{t'_2(s_2)^2(s-s_2)^2}+
  \frac{t'_4(s_2)}{t'_2(s_2)^2(s-s_2)}+
  \frac1{t'(s_A)(s-s_A)},
\end{equation}
 where, in the pole contributions, we
have used that
\begin{equation}
  -\frac{t''(s_A)}{2t'(s_A)^2}+ \frac{t''_4(s_2)}{2t'_2(s_2)^2}
=O(f^{-2}),
\label{eq:Aequal2}
\end{equation}
which, once again, can be safely neglected
since it correspond to the chiral expansion at very low energies.

Finally, if we evaluate perturbatively
\begin{equation}
  \frac1{t'(s_A)}\simeq
\frac1{t'_2(s_2)}-\frac{t'_4(s_2)}{t'_2(s_2)^2}+O(f^{-2}),
\label{eq:finallyA}
\end{equation}
and add 0 to Eq.\eqref{preIAM-A} written as
$1/t_2(s)-1/(t'_2(s_2)(s-s_2))$, we obtain for $t(s)$
Eq.\eqref{IAMext} with $A(s)$ given by Eq.\eqref{Aangelpipi}, i.e,
we recover our naive derivation of the modified IAM by subtracting
at the Adler zeros.

Note however, that in contrast to the
derivation where we used an arbitrary
$z_0$, now all subtractions
have been performed in the very low
energy region so that chiral expansions
for pole terms are well justified within ChPT.
The price to pay is that in the right cut integral terms
we have approximated $s_A$ by $s_2$, which is irrelevant in the
Adler zero region since the pole contributions dominate there,
and a remarkably good approximation
in the physical and resonance regions. Of course, we have still
used exact elastic unitarity in the
{\it integrands} over the physical cut,
which ensures that the modified
IAM satisfies exact elastic unitarity.

\section{Modified IAM:  unequal masses.} \label{sec:uneq}

When dealing with unequal masses, as in the $\pi K$ scattering case
that we will use for reference, in addition to the left and right
cuts, there is also a circular cut centered at $s=0$ with radius
$\sqrt{m_K^2-m_\pi^2}$ that contributes to the dispersion relation. 
This circular cut lies in the low energy region 
within the applicability range of ChPT and thus, 
as we do with the left cut, we will approximate the
inverse amplitude by its ChPT series to fourth order.
 Taking into account that $t_2(s)$ has no cuts we
will have $CC(1/t)=-CC(t_4/t_2^2)$, where $CC$ stands for the
circular cut contribution.  This is the same approximation used
before for the left cut and therefore we obtain  the same IAM
dispersive derivation. Hence, there is still the same problem with
Adler zeros and spurious poles.

The solution given in previous sections works similarly well for the 
 $I=3/2$ and $J=0$ partial wave.
 However, for the
$I=1/2$, $J=0$ $\pi K$ scattering channel complications arise due
to the form of the LO partial wave $t_2(s)$ which has
\emph{two} zeros instead of one.
\begin{equation}
  \label{eq:t2pik}
  t_2(s)=-\frac{5(s-s_{2+})(s-s_{2-})}{128\,\pi f s},
\end{equation}
with
$s_{2\pm}=\frac1{5}\left(m_K^2+m_\pi^2\pm
2\sqrt{4m_K^4-7m_K^2m_\pi^2+4m_\pi^4}\right)$
whose values are $s_{2+}=0.24\;{\rm GeV}^2$,
and $s_{2-}=-0.13\;{\rm GeV}^2$.
In particular, this means that, contrary to the previous cases $t''(s_{2\pm})\neq0$,
which complicates the derivation of a modified IAM.
Nevertheless, once again we have found for this special case of the
$I=1/2$, $J=0$ channel, a naive and two dispersive derivations,
that we detail next.

\subsection{Naive derivation, $I=1/2$, $J=0$ channel}

Following section 3, we define $1/t^{mIAM}=1/t^{IAM}+A^{mIAM}(s)/t_2^2(s)$ with
$A^{mIAM}(s)$  an analytic function at least off the real axis and
real for real $s$ to preserve the unitarity and
analytic properties of the original amplitude. Next we expand $1/t^{IAM}$
in Laurent series around $s=s_{2+}$, taking into account
that now $t''_2(s)\neq 0$, obtaining
\begin{equation}
  \label{tIAMexpansionpik}
  \frac1{t^{IAM}(s)}=-\frac{t_4(s_2)}{t'_2(s_{2+})^2(s-s_{2+})^2}+
  \frac{t'_2(s_{2+})-t'_4(s_{2+})}{t'_2(s_{2+})^2(s-s_{2+})}+
  \frac{t_4(s_{2+})t''_2(s_{2+})}{t'_2(s_{2+})^3(s-s_{2+})}+O(s-s_{2+})^0.
\end{equation}
As in section 3 we subtract the pole at $s_{2+}$ and add a pole at
$s_A$ to the inverse amplitude, and the constant in
the $s_A$ pole term is calculated perturbatively in order
to match the chiral expansion at low energies.
 Note that, following our previous
arguments,  we do not need to subtract the $s_{2-}$ pole, since $\im
t_4\neq 0$ on the LC, so that no spurious pole will appear in that
region. Proceeding then as in section \ref{sec:naive}, we get now:
\begin{equation}
  A^{mIAM}(s)= \frac{t_2(s)^2}{t_2'(s_{2+})^2}
  \left[
    \frac{t_4(s_{2+})}{(s-s_{2+})^2}-
    \frac{(s_{2+}-s_A)}{(s-s_{2+})(s-s_A)}
    \left(
      t'_2(s_{2+})-t_4'(s_{2+})+
      \frac{t_4(s_{2+})t''_2(s_{2+})}{t'_2(s_{2+})}
    \right)
  \right].
\label{Aangelpik}
\end{equation}

Then, Eq.\eqref{IAMext} with $A^{mIAM}(s)$ above unitarizes the
$I=1/2,\,J=0$ channel and has an Adler zero of the correct chiral order and has no
spurious pole. Let us remark that the above $A^{mIAM}(s)$ is a generalization also valid for
the equal mass case,  since it
is reduced to the $A^{mIAM}(s)$ we already obtained in Eq.\eqref{Aangelpipi} when $t_2''(s_{2+})=0$.

Apart from this naive formal derivation we can follow a
dispersive approach, detailed  next, in which we make use of analyticity and the
ChPT series is only used within its applicability region.

\subsection{Dispersive derivation at $z_0$, $I=1/2$, $J=0$ channel}

The derivation follows exactly that of section 2.1 but now, due to
the additional zero, $s_{2-}$, of $t_2$, which lies on the
negative axis, we cannot simply write, as usually done,
$\int_{LC}\disc t_4/t_2^2=2i\int_{LC}\im t_4/t_2^2$, nor
$\int_{LC}\disc 1/t_2=0$, where $\disc
f=f(x+i\epsilon)-f(x-i\epsilon)$ with real $x$. This zero also
modifies the form of the dispersion relation for $1/t_2$, that now
reads
\begin{equation}
  \label{1/t2disppik}
  \frac1{t_2(s)}=\frac1{t_2(z_0)}+PC_{+}(1/t_2)+PC_-(1/t_2),
  \quad PC_\pm(1/t_2)=
  \frac1{t'_2(s_{2\pm})}\left(\frac1{s-s_{2\pm}}-\frac1{z_0-s_{2\pm}} \right)
\end{equation}
Following the same steps as in section 2.1, we approximate $1/t$ on the left
cut by its chiral expansion, but now taking into account that, as
mentioned above, $\int_{LC}\,dz\,\disc\,(1/t_2(z))\neq 0$, i. e.
\begin{equation}
  \label{eq:disc1/t}
  LC(1/t)=\frac{s-z_0}{2\pi i}
  \int_{LC}\frac{\disc 1/t(z)}{(z-s)(z-z_0)}
  \simeq\frac{s-z_0}{2\pi i}\int_{LC}
  \frac{\disc (1/t_2(z)-t_4(z)/t_2(z)^2)}{(z-s)(z-z_0)},
\end{equation}
where we now get $-LC(t_4/t_2^2)$ as before, and a
new term coming from $1/t_2$,
  \begin{eqnarray}
    \label{eq:disc1/t2}
    \frac{s-z_0}{2\pi i}
    \int_{LC}\frac{\disc 1/t_2(z)}{(z-s)(z-z_0)}&=&
    \frac{s-z_0}{2\pi i}\int_{LC}
    \frac{(1/t_2(z+i\epsilon)-1/t_2(z-i\epsilon))}{(z-s)(z-z_0)}\\
   &=&
      -(s-z_0)\int_{LC}
      \frac{(z-s_{2-})/t_2(z)}{(z-s)(z-z_0)}\,\delta(z-s_{2-})
    =\frac1{t'_2(s_{2-})}\left(\frac1{s-s_{2-}}-
      \frac1{z_0-s_{2-}} \right)
\nonumber
  \end{eqnarray}
which is equal to $PC_-(1/t_2)$, so they will cancel in the
expression for $1/t$. Thus we again obtain Eq.\eqref{IAMpc}, but
now the explicit expression for $PC(t_4/t_2^2)$ at $s=s_{2+}$ is
different from Eq.\eqref{eq:PCt4/t2} because $t''_2(s)\neq 0$,
\begin{eqnarray}
  \label{PC0pik}
  PC(t_4/t_2^2)=\frac{t_4(s_{2+})}{t'_2(s_{2+})^2}
  \left(\frac1{(s-s_{2+})^2}-\frac1{(z_0-s_{2+})^2}\right)
  +\left(
    \frac{t'_4(s_{2+})}{t'_2(s_{2+})^2}-
    \frac{t_4(s_{2+})t''_2(s_{2+})}{t'_2(s_{2+})^3}
  \right)
  \left(
    \frac1{s-s_{2+}}-\frac1{z_0-s_{2+}}
    \right).
\end{eqnarray}
Finally, we get 
\begin{equation}
  \label{tmodpik}
  t^{z_0IAM}(s)=\frac{t_2(s)^2}{t_2(s)-t_4(s)+ A^{z_0IAM}(s)},
\end{equation}
with
\begin{equation}
\begin{aligned}
  \label{Apik}
   A^{z_0IAM}(s)&
  =t_2^2(s) \left(-PC_+(1/t_2)+PC(t_4/t_2^2)+PC(1/t)\right),
\end{aligned}
\end{equation}
where, again, we evaluate $1/t'(s_A)$ in $PC(1/t)$ perturbatively,
obtaining for $A (s)$
\begin{equation}
  \label{Az0pik}
  A^{z_0IAM}(s)=A^{mIAM}(s)-\frac{t_2(s)^2}{t_2(z_0)^2}A^{mIAM}(z_0)
\end{equation}
Formally, this looks the same as Eq.\eqref{Az0pipi} but now $A^{mIAM}(s)$
is of a more general form. Indeed, we obtain again
$A^{mIAM}(s)$  plus a term depending on $z_0$, which is also $O(f^{-6})$.
Nevertheless we will see that, as long as
$z_0$ remains in the range of validity of the approximations,  
the latter is numerically small not 
only in the physical region but also in the subthreshold region,
so that it
can be neglected to obtain the naive simpler formula, thus justifying numerically
the mIAM and the IAM.

However, a 
derivation of the mIAM that differs from the standard one 
since subtractions are 
made at the Adler zeros is also possible for the unequal mass case, 
and we detail it next.

\subsection{Dispersive derivation subtracting at the Adler zeros,
$I=1/2$, $J=0$ channel}

 The derivation follows closely that in section
\ref{subsAdlerpipi}, but now we will have extra terms (not
negligible in the chiral expansion) coming from $\disc 1/t_2$ in the
left cut when $s_A$ is expanded around $s_2$. In addition, the
expressions for the pole contributions are more complicated because
$t''_2(s)\neq 0$.

We again expand the left cut to NLO:
\begin{eqnarray}
  \label{LCpikexp}
  LC(1/t)
  &\simeq& \frac{s-s_{2+}-s_4}{2\pi i}\int_{LC}dz\,
  \frac{\disc (1/t_2(z)-t_4(z)/t_2(z)^2)}{(z-s)(z-s_{2+})}\left(
    1+\frac{s_4}{z-s_{2+}}\right)\\\nonumber
  &\simeq& -LC(t_4/t_2^2)+
  \frac{s-s_{2+}}{t'(s_{2-})(s-s_{2-})(s_{2-}-s_{2+})}-
  \frac{t_4(s_{2+})}{t'_2(s_{2+})t'_2(s_{2-})(s_{2+}-s_{2-})^2},
\end{eqnarray}
where we obtain $-LC(t_4/t_2^2)$ plus two extra terms.
The pole contributions now read
\begin{eqnarray}
  \label{pc1/tpikAd}
    PC(1/t)&=&\frac1{t'(s_A)(s-s_A)}-\frac{t''(s_A)}{2t'(s_A)^2},\\
  \label{PCt4/t2spikAd}
  PC(t_4/t_2^2)&=&\frac{t_4(s_{2+})}{t'_2(s_{2+})^2(s-s_{2+})^2}+
  \left(
    \frac{t'_4(s_{2+})}{t'_2(s_{2+})^2}-
    \frac{t_4(s_{2+})t''_2(s_{2+})}{t'_2(s_{2+})^3}
  \right)\frac1{s-s_{2+}}\\\nonumber
  &+&\frac{t''_4(s_{2+})}{2t'_2(s_{2+})^2}-
  \frac{t'_4(s_{2+})t''_2(s_{2+})}{t'_2(s_{2+})^3}+
  \frac{3t_4(s_{2+})t''_2(s_{2+})^2}{4t'_2(s_{2+})^4}-
  \frac{t_4(s_{2+})t'''_2(s_{2+})}{3t'_2(s_{2+})^3}.
\end{eqnarray}
where all terms in $PC(1/t)$ will be evaluated perturbatively
excepting $1/(s-s_A)$, that is the one that gives the needed pole at
$s_A$ to the inverse amplitude. Then, adding the dispersion
relations for $1/t$ and $t_4/t_2^2$, and using that
\begin{equation}
  \label{1/t2pikAd}
  \frac1{t_2(s)}=\frac1{t'_2(s_{2+})(s-s_{2+})}+
  \frac{s-s_{2+}}{t'_2(s_{2-})(s-s_{2-})(s_{2-}-s_{2+})}-
  \frac{t''_2(s_{2+})}{2t'_2(s_{2+})^2}
\end{equation}
we thus arrive to $A^{mIAM}$ in Eq.\eqref{Aangelpik} plus $O(f^{-6})$ terms that
can be safely neglected since they all correspond to the
chiral expansion at very low energies.

\section{Numerical Results: comparison between  
different approaches and $z_0$ sensitivity.}
\label{sec:num}

In this section we compare numerically the IAM with the modified methods we have
derived. The precise values of the ChPT low energy constants are not relevant
for what we want to show here. Just for illustration, we 
taken for SU(2) ChPT the typical values of $l_3^r=0.82,\,l_4^r=6.2$ from
the second reference in \cite{ChPT},
at a renormalization scale $\mu=770$ MeV.
The values 
$l_1^r=-3.7,\,l_2^r=5.0$ have been obtained from an IAM fit to
$\pi\pi$ scattering data. In particular, 
we have used the same data sets used in the IAM fits of the second reference 
in \cite{Guerrero:1998ei} but
we have updated
the controversial $f_0(600)$ channel by using the choice of data 
explained in \cite{Yndurain:2007qm} consistent with
forward dispersion relations and Roy equations as shown in \cite{Kaminski:2006qe}.
For $\pi K$ scattering we have chosen 
the central values of the 
$SU(3)$ ChPT constants of the IAM I set in the last reference 
of \cite{Guerrero:1998ei}. 

\begin{figure}
  \centering
  \vbox{
    \hbox{
     \includegraphics[scale=.45]{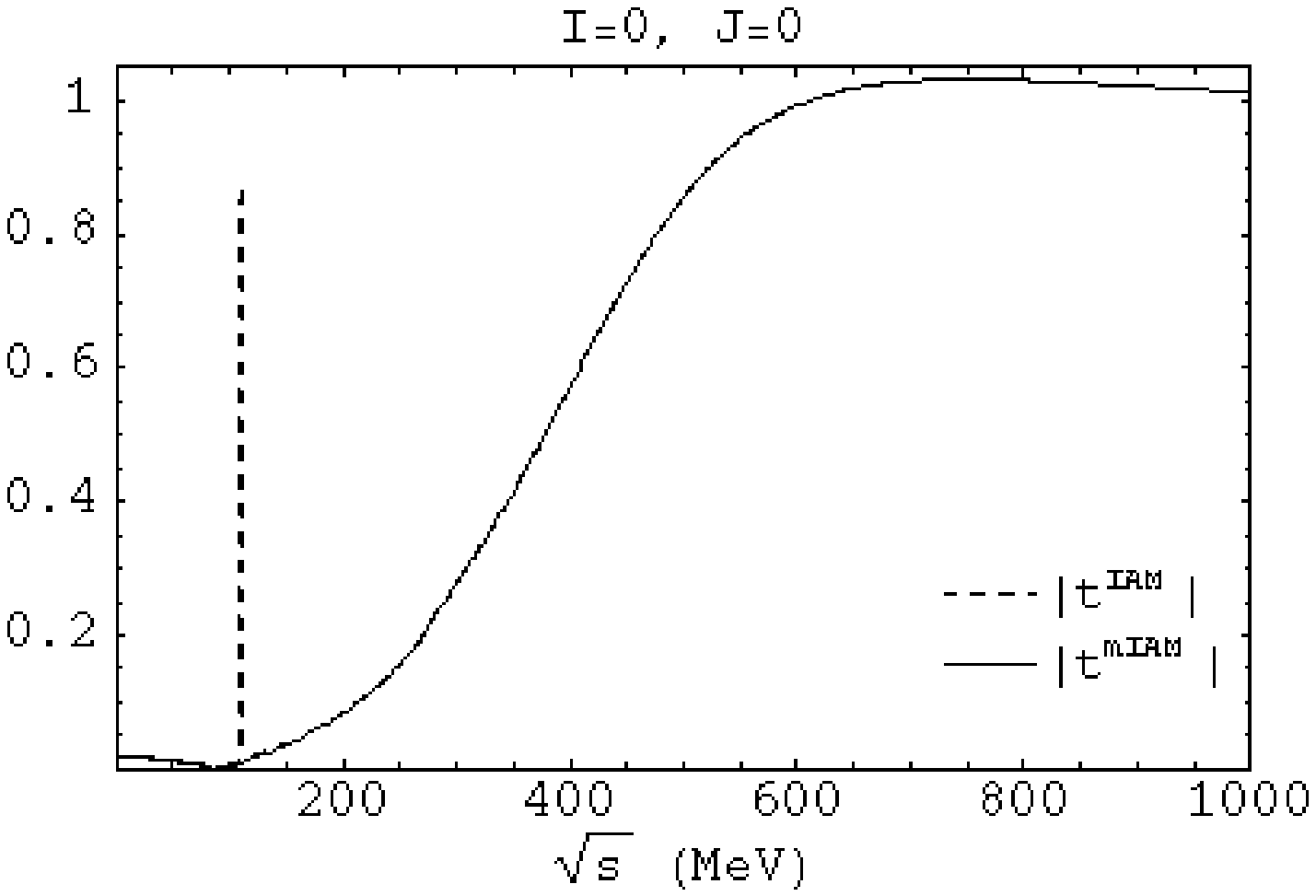}
     \includegraphics[scale=.45]{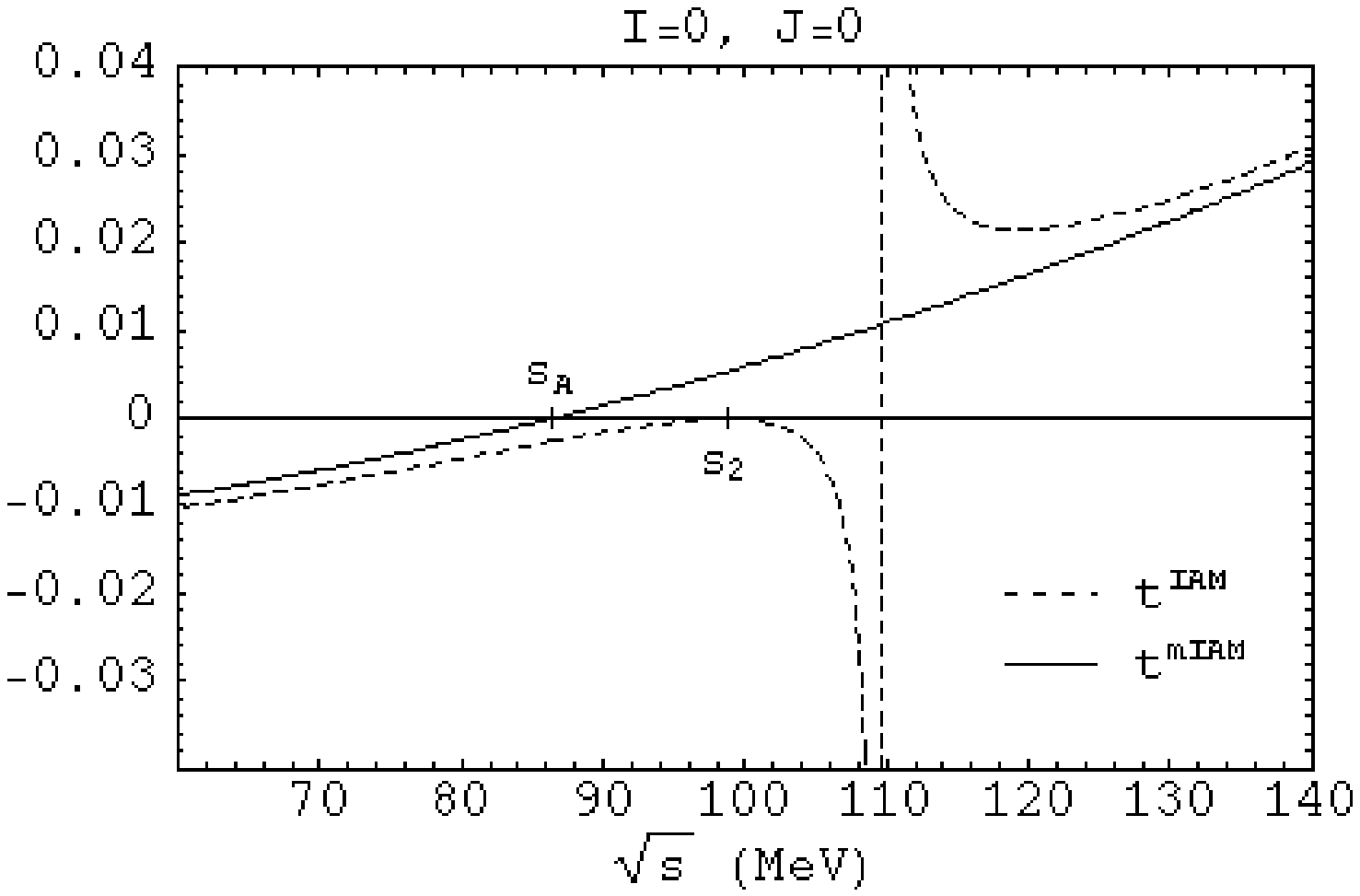}
    }
    \hbox{
     \includegraphics[scale=.45]{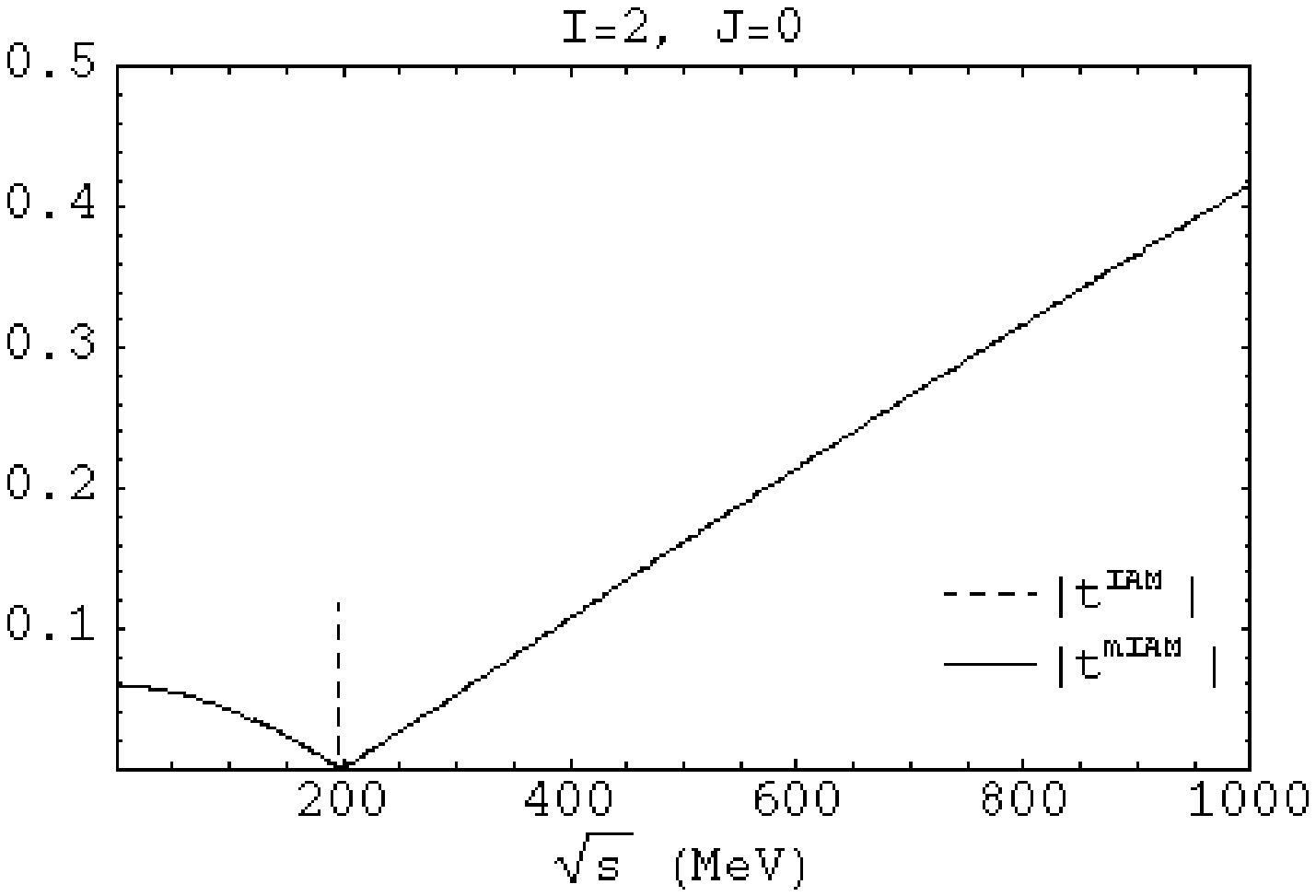}
     \includegraphics[scale=.45]{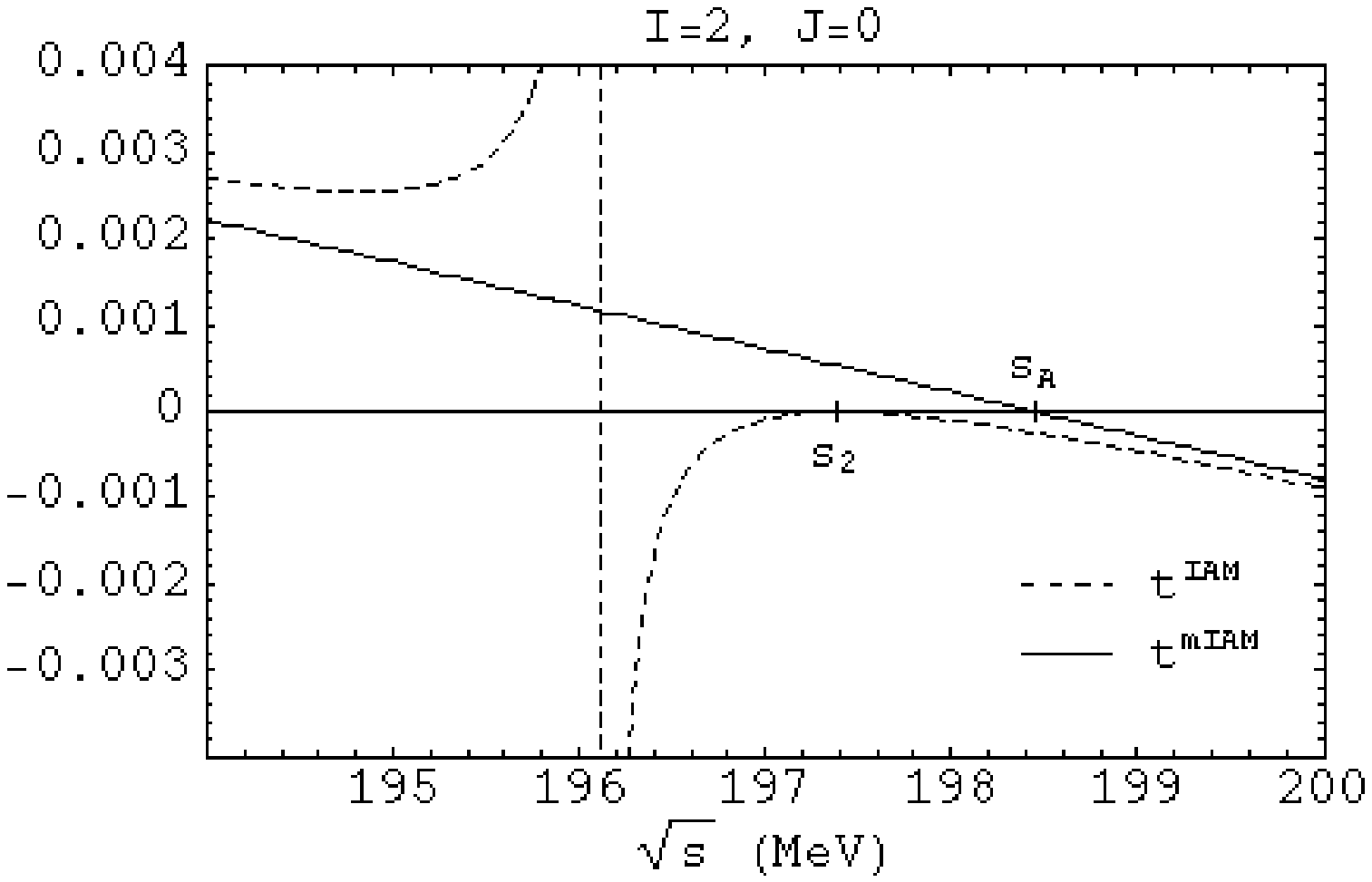}
    }
    \hbox{
     \includegraphics[scale=.45]{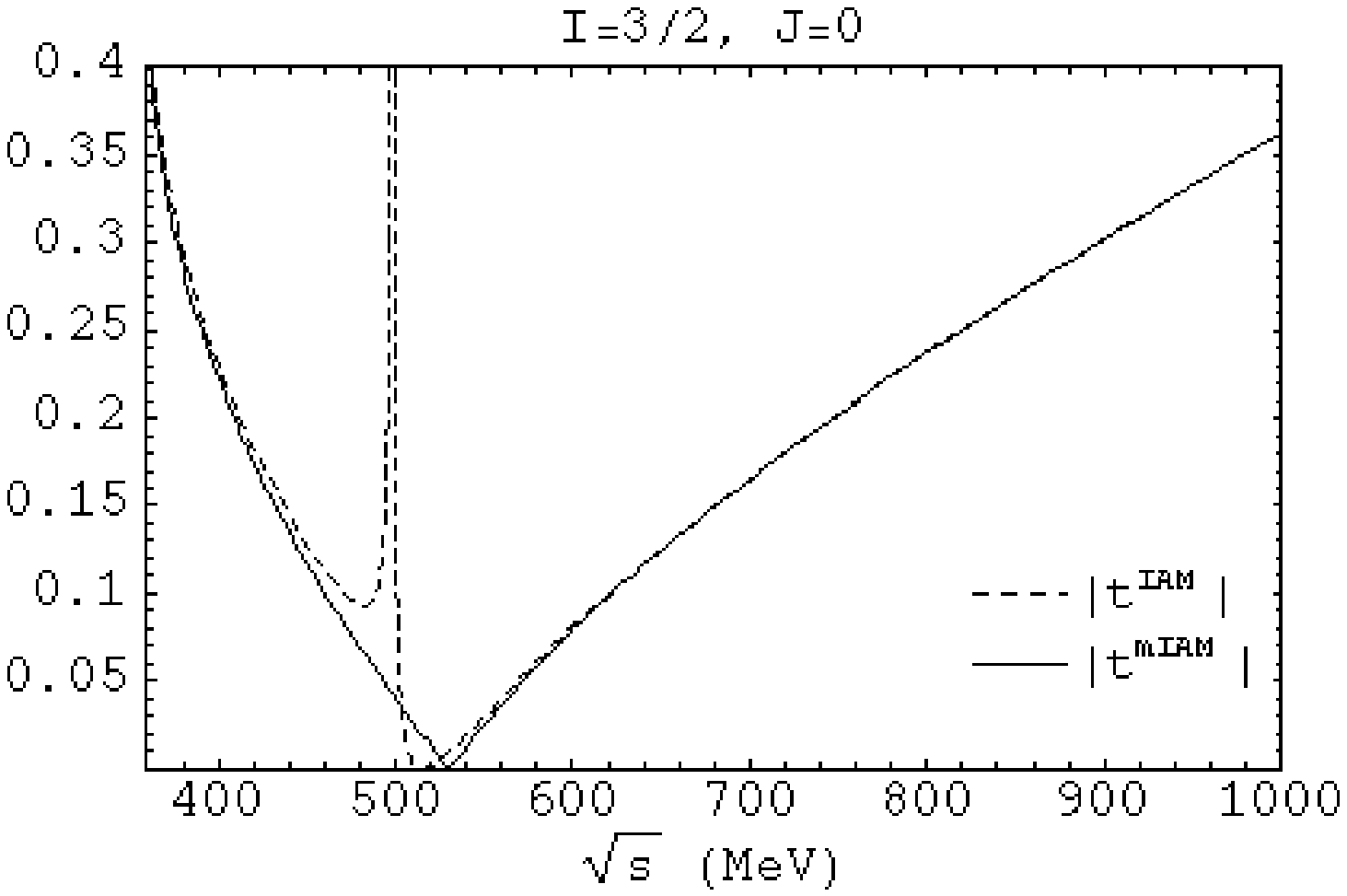}
     \includegraphics[scale=.45]{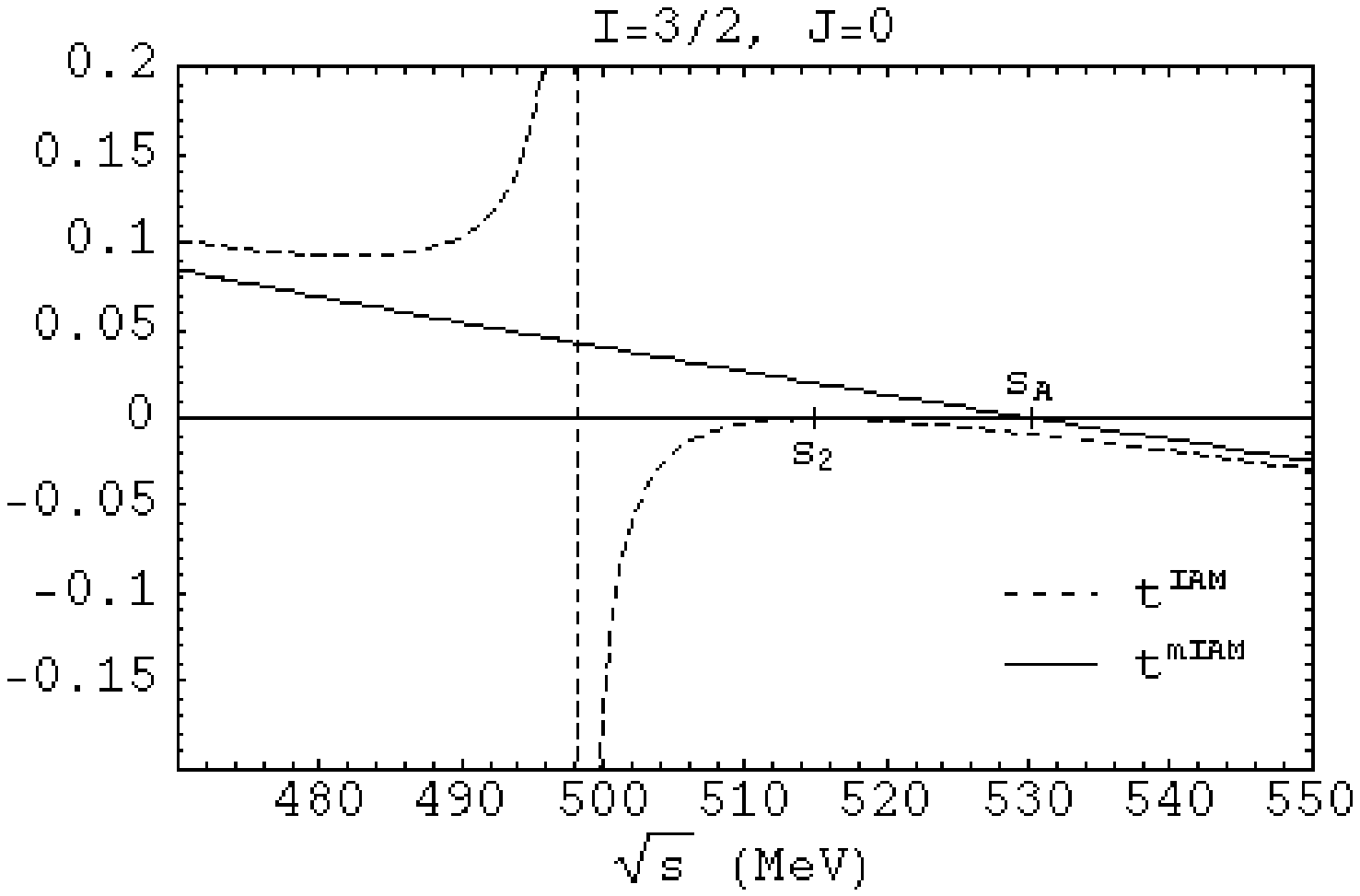}
    }
    \hbox{
     \includegraphics[scale=.45]{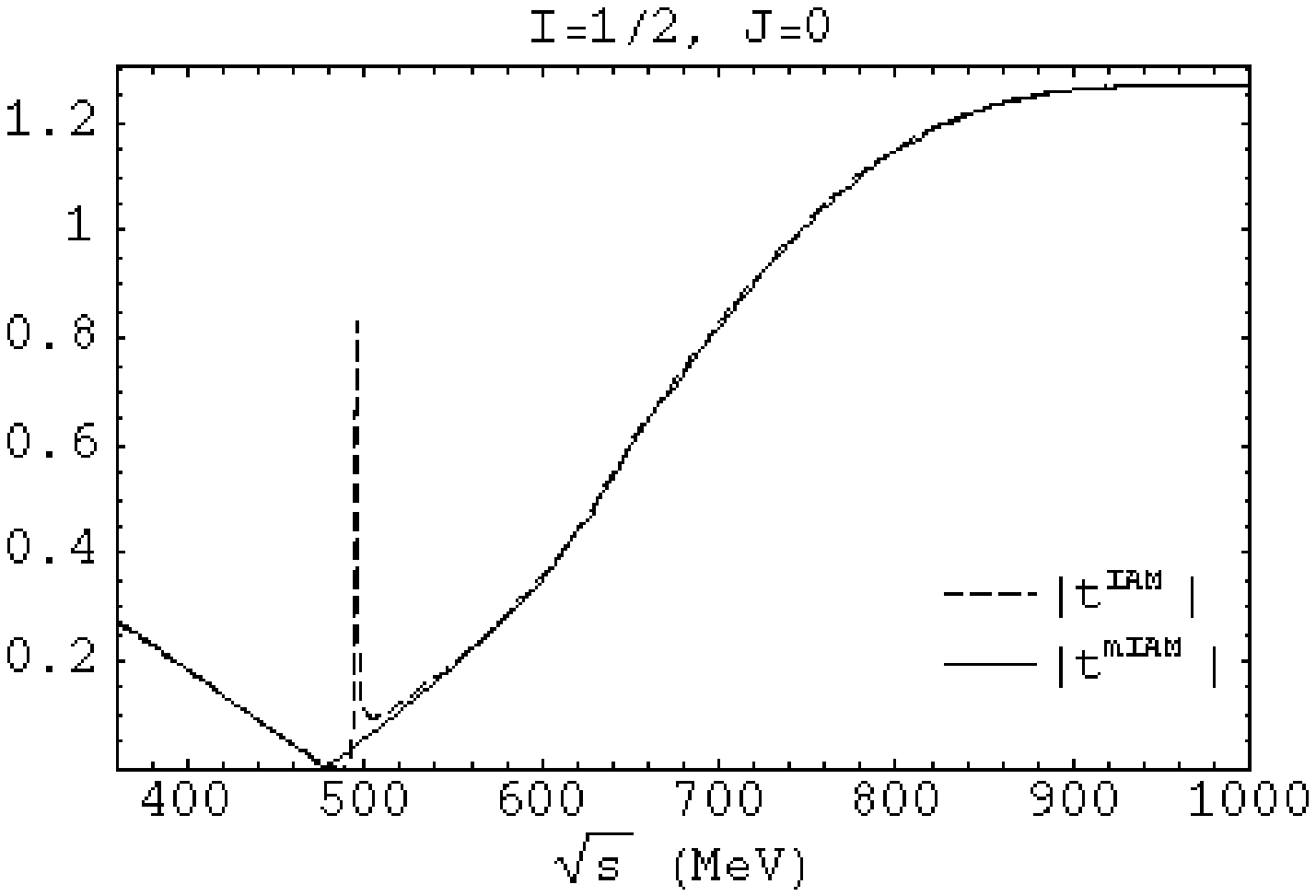}
     \includegraphics[scale=.45]{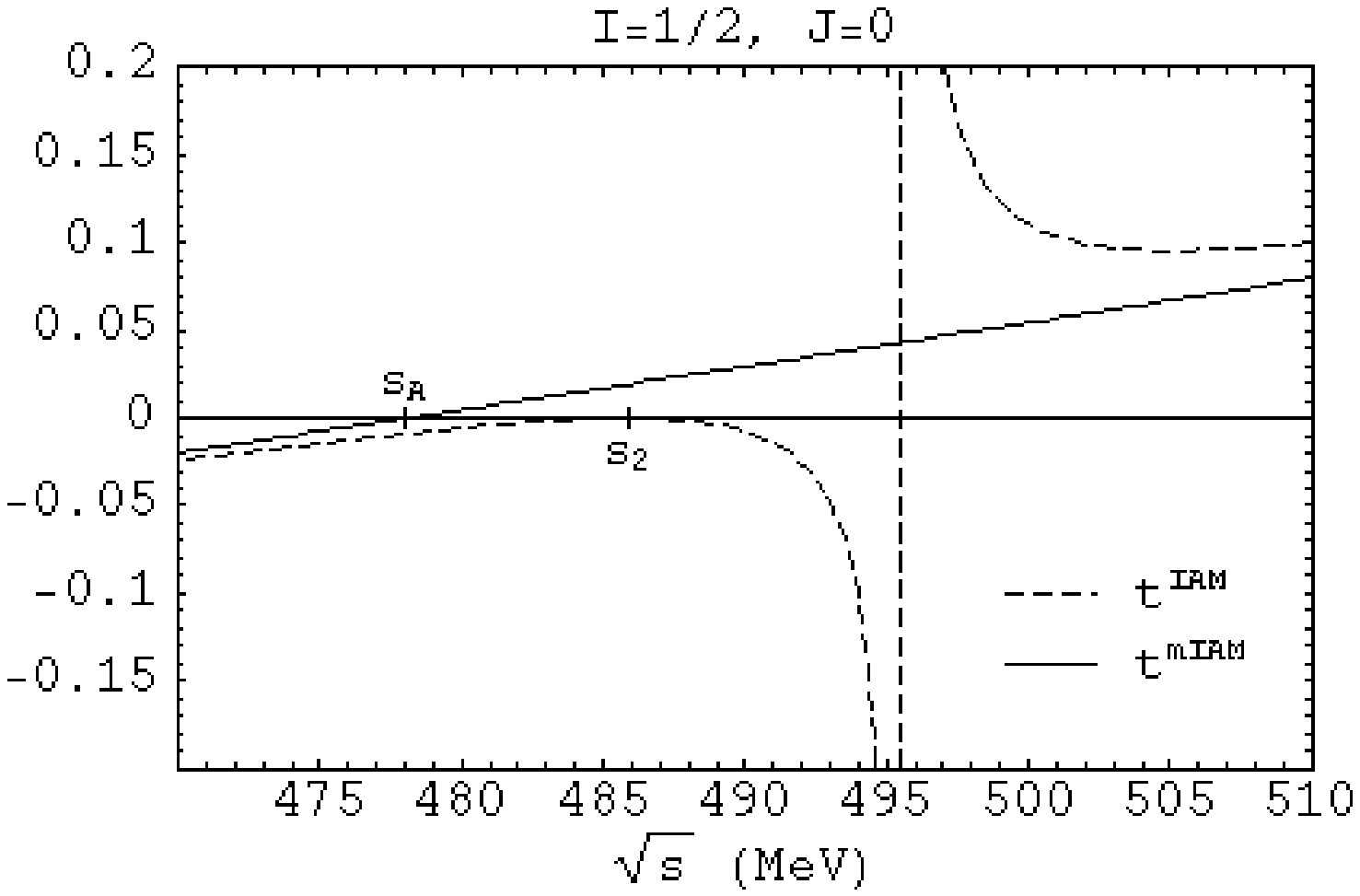}
    }
  }
  \caption{\label{fig:comparison}
   Comparison between the IAM and the
  mIAM for different isospin  $I$ partial waves of $\pi\pi$ and $\pi K$ in the scalar $J=0$ channel. The left column covers 
the region from the left cut up to 1 GeV. The only significant differences between both methods occurs in the region around the Adler 
zeros of each partial wave, which is shown in detail in the right column.}
\end{figure}

We show in Fig.\ref{fig:comparison} several plots comparing the IAM
and mIAM results for the modulus of different partial 
waves of $\pi\pi$ and$\pi K$ elastic scattering in the 
scalar channel. We see that both methods are indistinguishable in the 
physical region, shown in the left column, and only differ in the Adler
zero region which is shown in detail in the right column. Note that the IAM
has a spurious pole and a double zero in that region, whereas the mIAM
does not have such pole and its zero is simple.

We have also calculated the $\sigma$ and $\kappa$
pole positions with the IAM and the mIAM, obtaining the
same pole position within $\sim 1\, {\rm MeV}$, as shown
in Table \ref{poles}.
\begin{table}[h]
  \centering
  \begin{tabular}{|c|c|c|}
    \hline
    Method        & $\sigma$ pole & $\kappa$ pole\\
    \hline
    IAM           & $443.71-i\,217.58$ & $724.2-i\,216.2$\\
    mIAM        & $443.68-i\,217.56$ & $725.3-i\,216.3$\\
    $z_0$IAM, $z_0=s_{th}$  & $443.82-i\,216.99$ & $727.7-i\,210.0$\\
    \hline
  \end{tabular}
  \caption{\label{poles} $\sigma$ and $\kappa$ pole positions calculated
    with the IAM, the mIAM and the $z_0$IAM with $z_0=s_{th}$}
\end{table}

\begin{figure}
\centering
  \vbox{
    \hbox{
     \includegraphics[scale=.45]{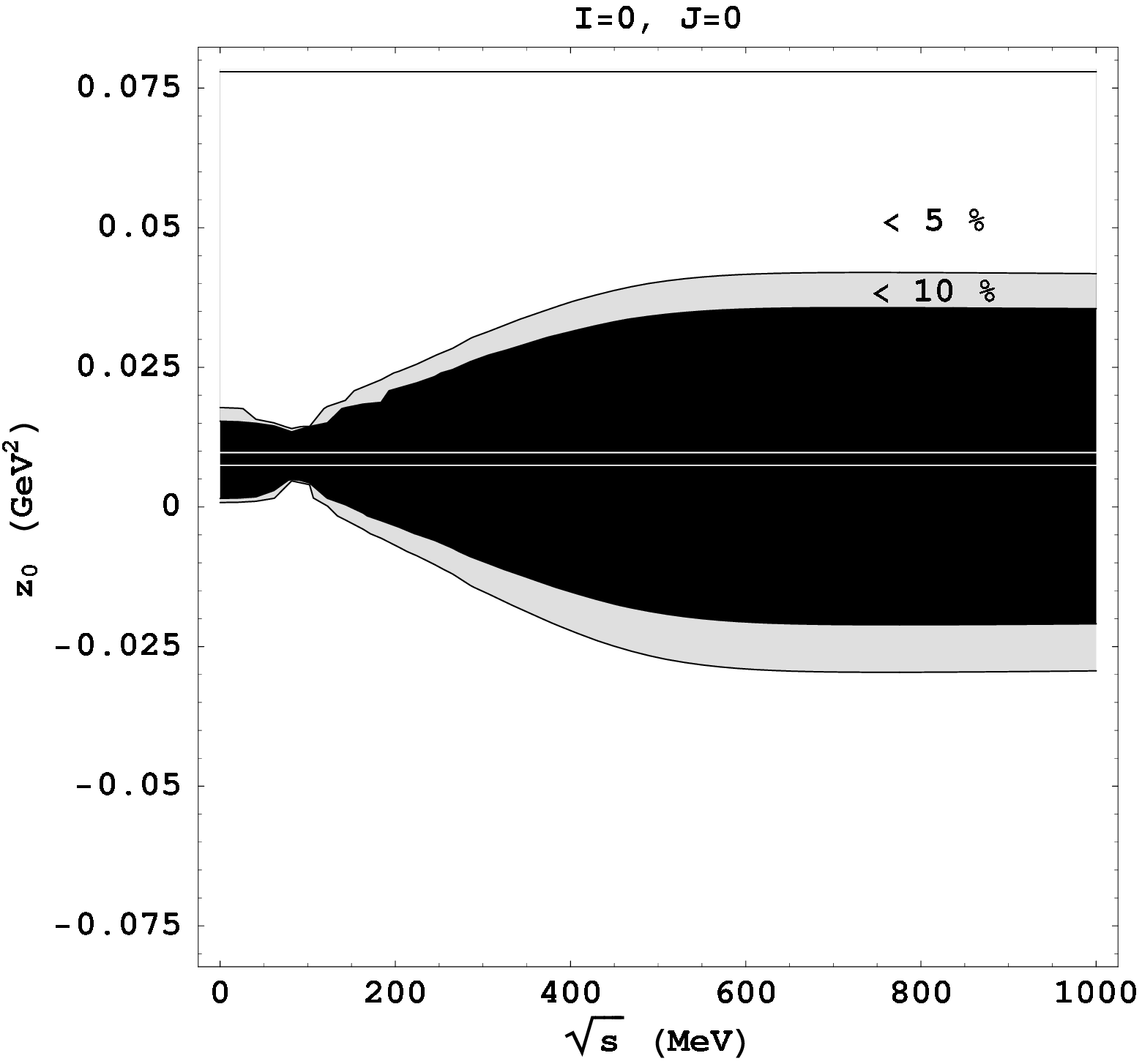}
     \includegraphics[scale=.45]{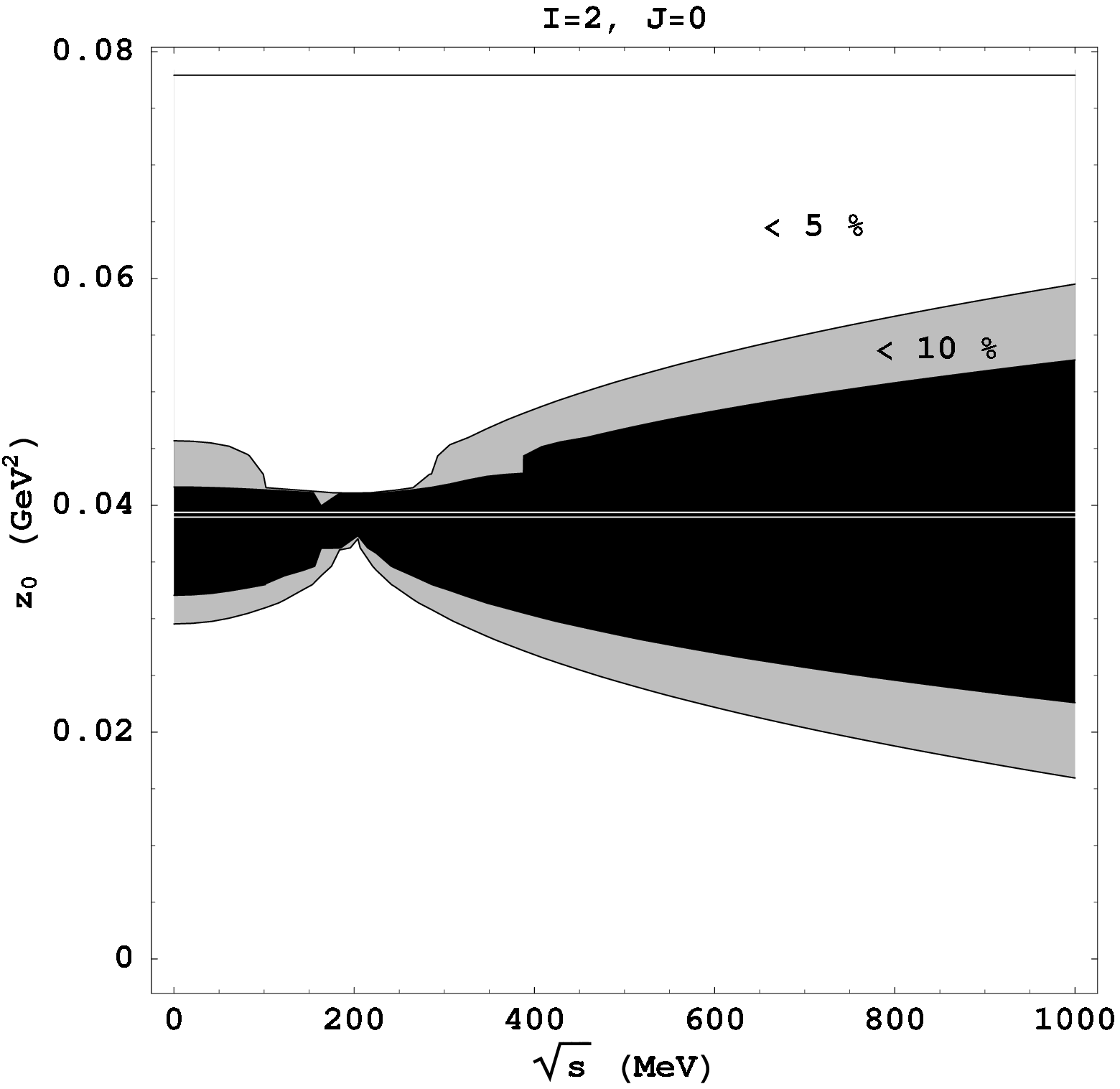}
    }
    \hbox{
     \includegraphics[scale=.45]{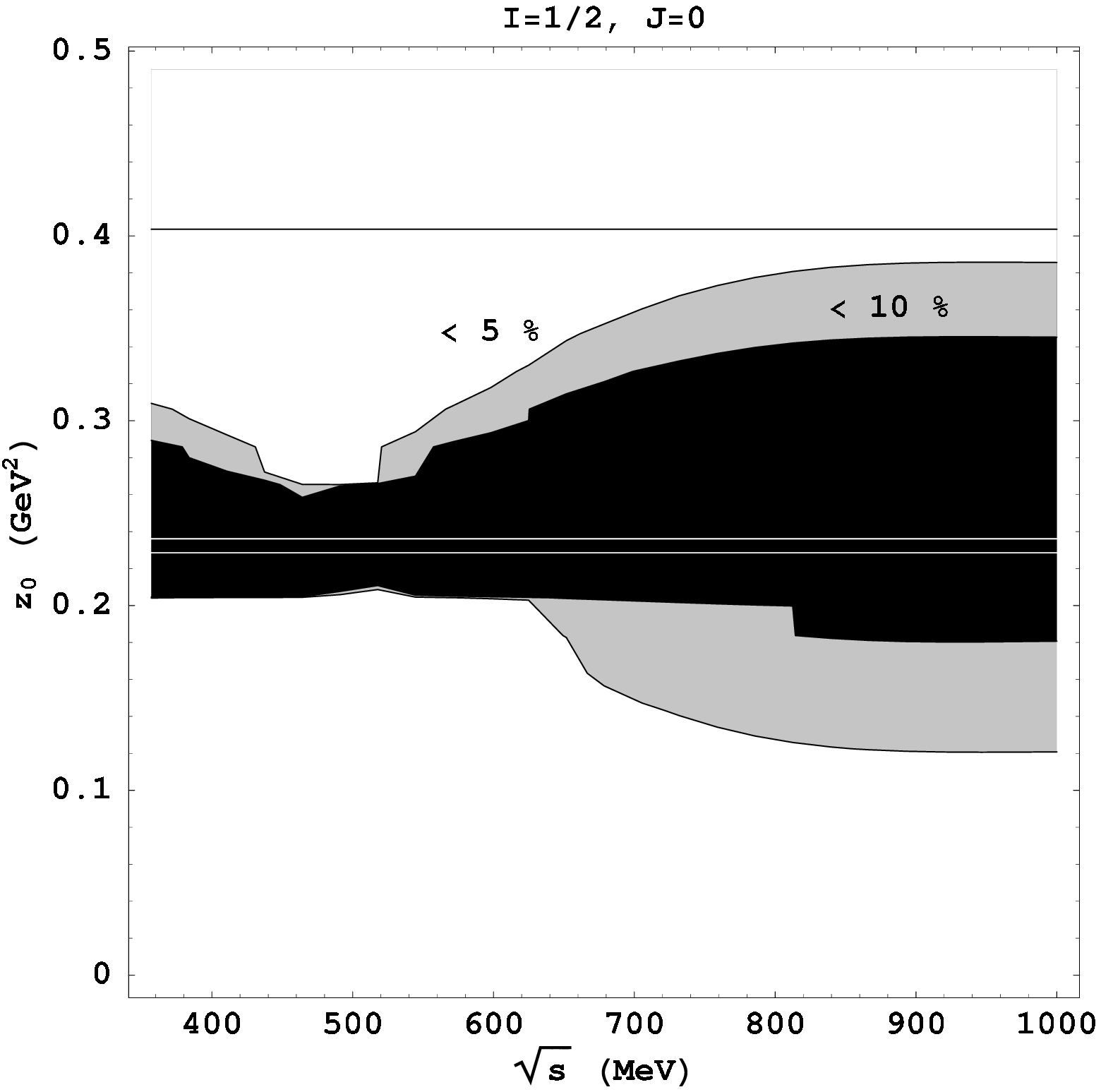}
     \includegraphics[scale=.45]{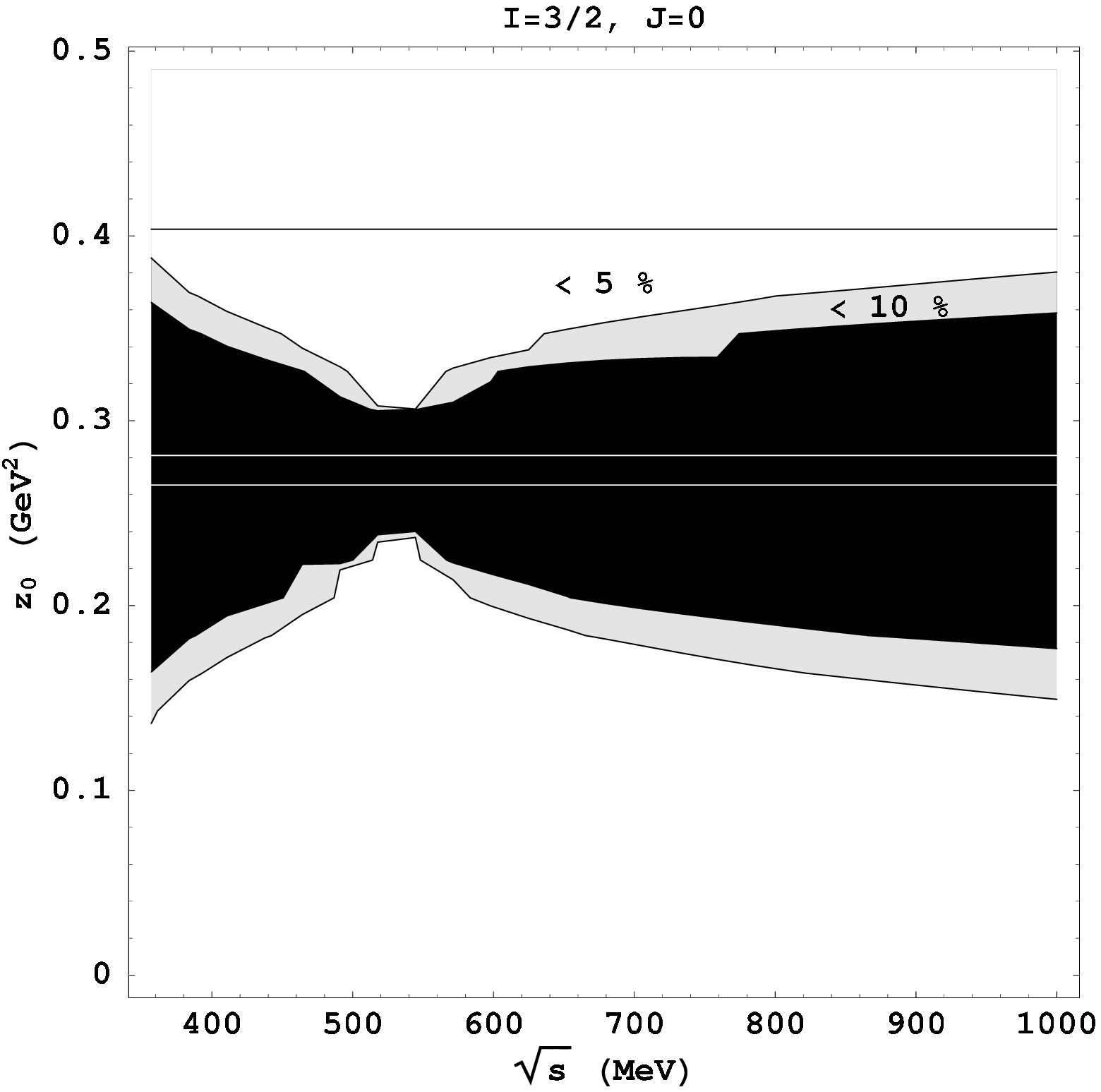}
    }
  }
  \caption{\label{contourplots} Contour plots of the relative differences
between the mIAM and the $z_0$IAM in the $(s,z_0)$ plane. 
We see that,  as long as $z_0$  lies in the low energy region but
sufficiently far from the the Adler zeros $s_A$ and $s_2$ (white lines) 
the differences become small for all partial waves. 
Note that for $z_0=s_{th}$ (black line) the differences for
  all cases are less than 5\%.}
\end{figure}

In the derivation of the $z_0$IAM, we have an arbitrary subtraction point
$z_0\neq s_2,s_A$, but
we only know how to calculate the amplitude at $z_0$ if we can use the chiral expansion
$1/t(z_0)\simeq 1/t_2(z_0)-t_4(z_0)/t_2(z_0)^2$, which is only
valid if $z_0$ lies on the low energy region. Also, due to
$t_2$ having a zero at $s_2$, the above expansion is a very bad approximation
if $z_0$ is near $s_2$. We can estimate how close $z_0$ 
could be from $s_A$ and $s_2$
by looking at the expansion
\begin{equation}
  \label{z0exp}\frac1{t(z_0)}\simeq
  \frac1{t_2(z_0)}-\frac{t_4(z_0)}{t_2(z_0)^2}\simeq
  \frac1{t_2(z_0)}\left(
    1+\frac{s_4}{(z_0-s_2)}+\cdots
  \right).
\end{equation}
where we have only made explicit the $s_4/(z_0-s_2)$ pole term.
Hence, for our approximations to remain valid, we have to
make sure that the ratio $s_4/(z_0-s_2)$ is small enough. 
Thus, we show in Fig.\ref{contourplots} the contour plots 
in the energy and $z_0$ plane of the 
relative difference between the mIAM and the $z_0$IAM
\begin{equation}
  \label{diff}
  \Delta=\frac{|t^{mIAM}(s)-t^{z_0IAM}(s)|}
  {\frac1{2}|t^{mIAM}(s)+t^{z_0IAM}(s)|},
\end{equation}
as a function of the energy and $z_0$. We show two contour lines
corresponding to $\Delta=10\%$ and $5\%$. We see that, as long
as the choice of $z_0$ is sufficiently far from $s_A$ and $s_2$
(the white lines in the plots)
the result of the $z_0$IAM differs little
from the mIAM. In particular, we have checked that
in order to obtain a relative difference less than 10\%  and 5\%, 
 in the worst case, which is the $I=2$, $J=0$ wave, 
we have to be sure that our choice of 
subtraction point $z_0$  makes the ratio of 
$|s_4/(z_0-s_2)|<3\%$
and $2\%$, respectively. 
Let us recall that an error of 5\% is a very precise result in this
context, since it is 
 much less than the uncertainties (mostly of systematic origin)
of the existing data on meson-meson scattering. 
For instance, just the isospin violation effects, 
that are usually not taken into account
by these experiments, can be estimated at the level of 2 or 3\%,
and these are added to further statistical and very big systematic
uncertainties.

In summary, the choice of subtraction constant has little relevance
for the $z_0$IAM, as long as it lies in a place 
where the NLO chiral expansion of the inverse amplitude 
is a good approximation to the inverse itself. In such case,
the $z_0$IAM results are very close to those of the mIAM, 
and therefore to those
of the standard IAM in the physical region.
This provides is a strong check of the stability and robustness 
of the standard NLO IAM results.

\begin{figure}[h]
    \centering
  \hbox{
    \includegraphics[scale=.53]{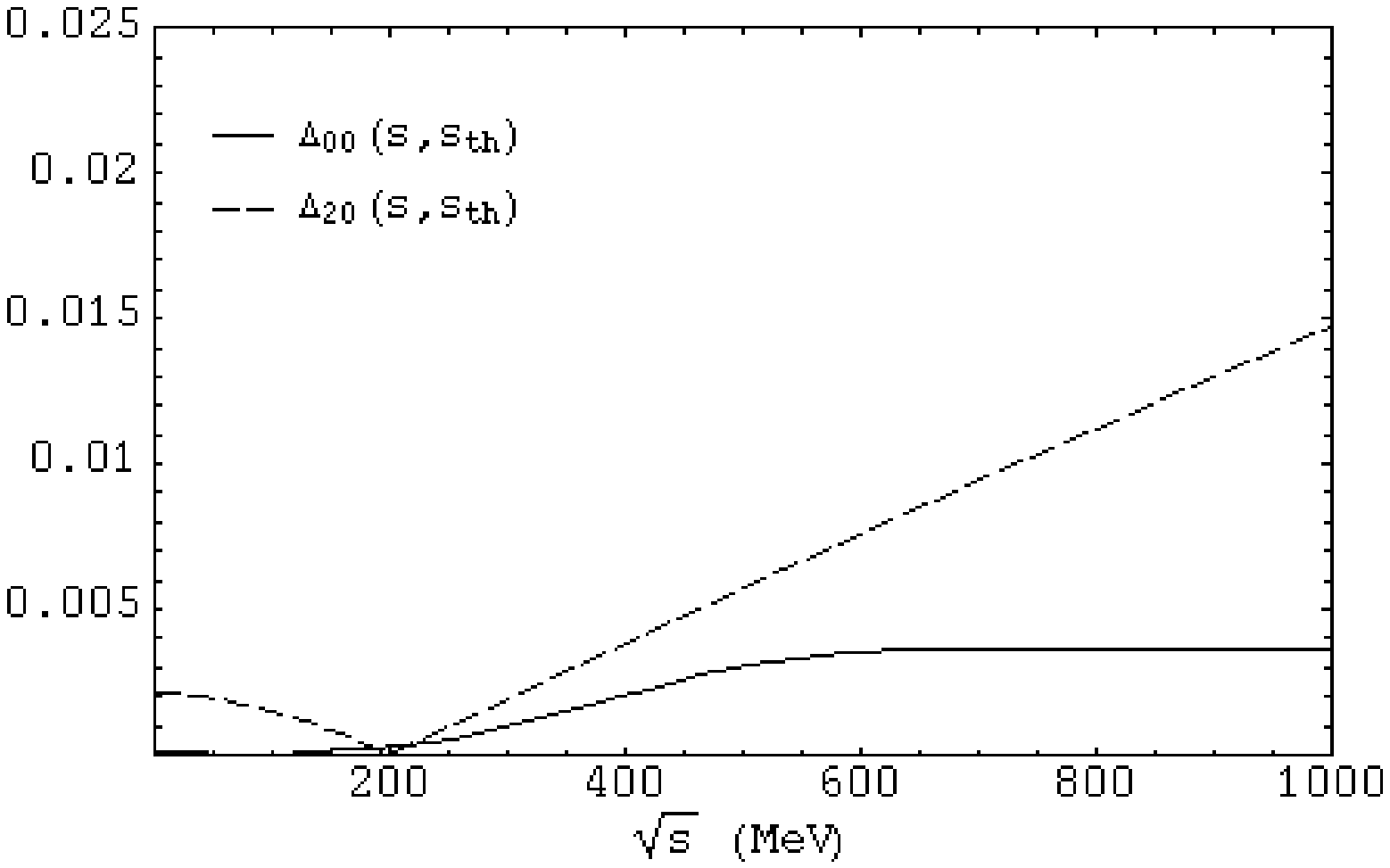}
    \includegraphics[scale=.53]{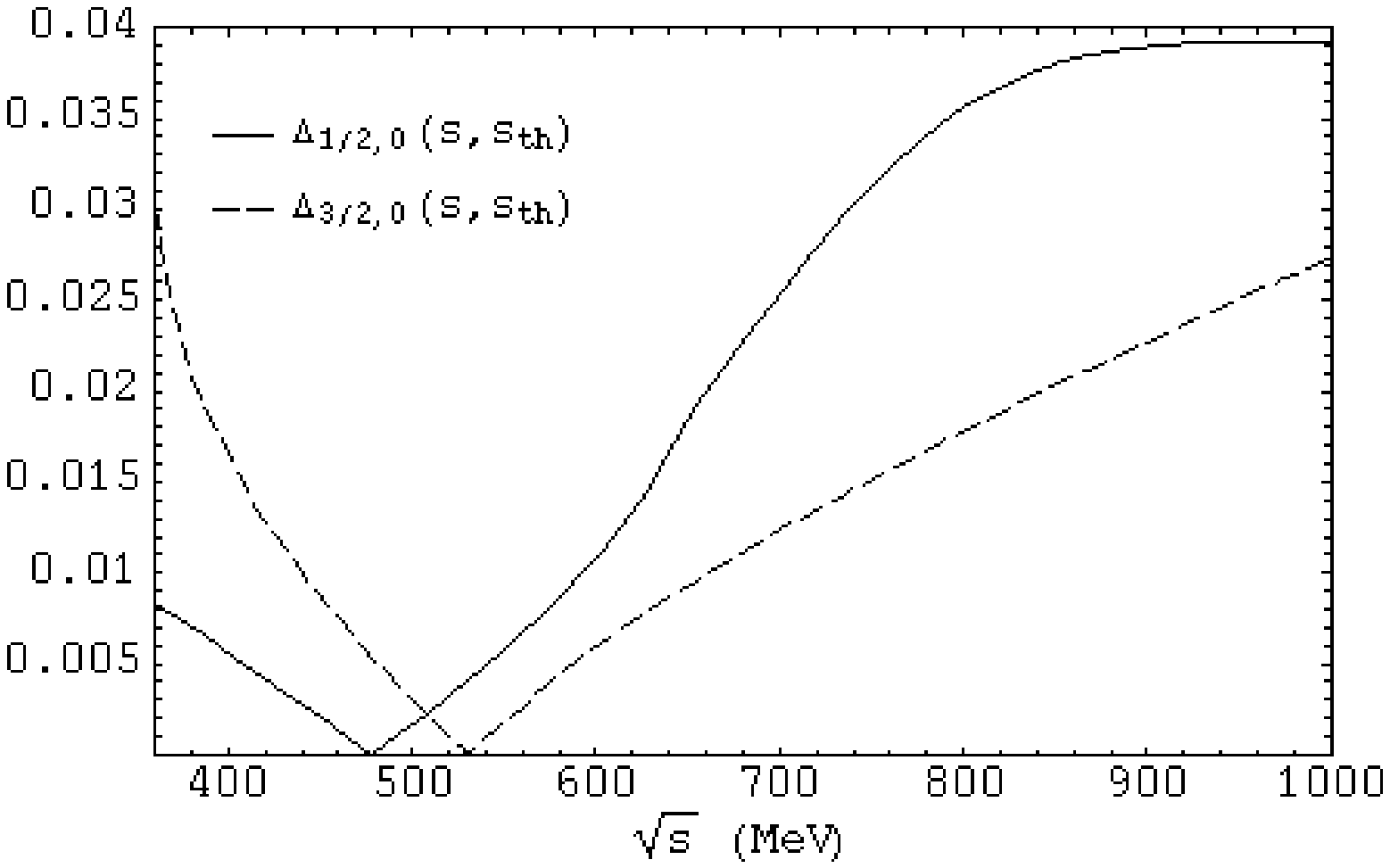}
  }
  \caption{\label{fig:diff} Relative differences $\Delta$, 
with $z_0=s_{th}$, for $\pi\pi$ (left), and $\pi K$ scattering (right). 
Note that the differences are less than 1.5\% and 4\%, respectively.}
\end{figure}

We have also shown in the  plots of Fig. 3 a line at $z_0=s_{th}$, 
which is a very natural choice of subtraction point, since the 
subtraction constants can then be written in terms of threshold parameters, which are well studied in the literature. We
see that, with this choice of $z_0$,
 in all cases we have a relative difference $\Delta$ which is less than 5\%.
Actually, it is even smaller, as seen
in Fig.\ref{fig:diff}, where we plot in detail the results for $\Delta$
subtracting at $z_0=s_{th}$, and we see that for $\pi\pi$ we have
a relative difference less than 1.5\% for all energies and less than
4\% for the $\pi K$ case.

\section{Discussion and summary}

In this work we have shown that it is possible to modify slightly
the one-channel Inverse Amplitude Method (IAM) so that, also in the
scalar channel, it provides a reliable description of the unitarized
partial wave amplitudes below threshold. In particular, we have
shown that it is possible to obtain a modified IAM that has the
Adler zeros located in the same place as the Effective Chiral
Expansion up to the desired order, that these zeros are single, and
that the spurious poles below threshold that occur in the standard IAM
are no longer present.

The IAM has been most frequently used at NLO,
where the chiral expansion of a partial wave is written as
$t(s)=t_2(s)+t_4(s)+...$. For such a case the simplest
modification to the IAM that we have found can be
 written as follows:
\begin{equation}
t^{mIAM}(s)= \frac{t^2_2(s)}{ t_2(s)-t_4 (s)+A^{mIAM}(s)},
\label{IAMfinal}
\end{equation}
where
\begin{equation}
  A^{mIAM}(s)= \frac{t_2(s)^2}{t_2'(s_2)^2}\left[\frac{t_4(s_2)}{(s-s_2)^2}-
\frac{(s_2-s_A)}{(s-s_2)(s-s_A)}
\left(t'_2(s_2)-t_4'(s_2)+\frac{t_4(s_2)t''_2(s_2)}{t'_2(s_2)}\right)
  \right]
\label{Afinal}
\end{equation}
and in order to have the Adler zero 
exactly on its NLO position, we set $s_A\rightarrow s_2+s_4$ where
$s_2=O(m^2)$ and $s_4\simeq O(m^4/f^2)$ are the Adler zeros at
LO and its NLO correction, respectively. That is, they are
obtained from $t_2(s_2)=0$ and $t_2(s_2+s_4)+t_4(s_2+s_4)=0$. In
general, $s_4$ should be calculated numerically. The above formula
is valid both for the elastic scattering of equal 
and unequal meson masses. However, in most of the cases, like $\pi\pi$ scattering,
the formula simplifies further since
 $t''_2(s_2)=0$
and $t_2(s)=t'_2(s_2)(s-s_2)$. Note also that the $A^{mIAM}(s)$ piece
counts as next to next to leading order (NNLO) in the chiral expansion,
and for that reason it was neglected in the standard IAM
derivation, which is recovered by setting $A^{mIAM}(s)\rightarrow0$ and remains
valid in the physical region, since there $A^{mIAM}(s)$ is indeed negligible.

In Sect.3 we have given  a ``naive'' formal derivation of 
$A^{mIAM}(s)$ by adding, in a rather {\it ad hoc}
way, the pieces needed to
fix the Adler zero and spurious pole problems without spoiling
unitarity and the chiral symmetry expansion. 
However, in Sects. 4 and 5 we have
shown that the mIAM formulae above can be derived by using 
the analytic properties of amplitudes in the form
of  dispersion relations and
imposing elastic unitarity on the right cut. The use of the Chiral
Effective Expansion is well justified to calculate the subtraction
constants and pole contributions to the dispersion relations, and is
also used to approximate contributions from other cuts. In
particular the integral over the left cut is calculated to NLO,
 which is a good approximation in the low energy
region that dominates the integral. Therefore, there are no model
dependencies but just approximations within the effective theory up
to a given order. This allows for a straightforward and systematic
extension of the elastic IAM and modified IAM to higher orders.

A usual criticism to unitarization methods is their arbitrariness,
but we have shown here that the IAM, modified or not,
 is not just unitarizing, but
also imposing a stringent analytic structure on the amplitudes, something
that leaves little room for such arbitrariness:
the choice of the subtraction points.
However, if the chiral effective expansion is to be used to calculate
the amplitude at the subtraction points or pole positions, the subtraction
points should lie, first of all, in the low energy region.
But, in addition, since a dispersion relation is written
for the inverse amplitude, that subtraction point should also
lie far from the Adler zero.
In this work, we have explicitly shown that
as long as those two constraints are satisfied, the choice of 
subtraction point 
has a very small numerical effect
on the resulting amplitude. Moreover, we have shown
that the results of the standard IAM in the physical axis
and resonance region in the complex plane remain unchanged
when using the modified IAM and are extremely stable under
different choices of subtraction points.

In summary, we have presented a slightly modified Inverse Amplitude
Method for the elastic case, that has the Adler zeros in the correct
position and of the correct order and no spurious poles in that
region. We have shown that the results already obtained  with the
standard IAM are robust in the physical region, where it can still
be used safely, but the new modifications allow for the study of the
subthreshold region that is of interest in problems like thermal
restoration of chiral symmetry \cite{FernandezFraile:2007fv}
or the quark mass dependence \cite{inprep} of
resonances in meson-meson scattering.

\section*{Acknowledgments}
Research partially funded by Banco Santander/Complutense contract
PR27/05-13955-BSCH and Spanish  contracts  FPA2007-29115-E,
FPA2005-02327, FPA2004-02602, UCM-CAM 910309. J.R.
Pel\'aez research is partly funded by  Spanish  contract BFM2003-00856,
and is part of the EU integrated infrastructure
initiative HADRONPHYSICS PROJECT, under contract
RII3-CT-2004-506078.

\appendix
\section{Chiral counting of pole contributions}

Next we will check by explicit calculation that the total pole
contribution counts as $\mo(f^{-6})$ in the amplitude. We will do it
for the more general case where $t''_2(s)\neq 0$. In
particular, we only need to expand $PC(1/t)$. First note that
\begin{equation}
  \label{PCexpansion}
  t'(s_A)=t'_2(s_2+s_4+\cdots)+
  t'_4(s_2+s_4+\cdots)+\mo (f^{-6})=t'_2(s_2)+t''_2(s_2)s_4+t'_4(s_2)+
  \mo (f^{-6}).
\end{equation}
With this
\begin{equation}
  \frac1{t'(s_A)}=\frac1{t'_2(s_2)}-
  \frac{t'_4(s_2)}{t'_2(s_2)^2}+
  \frac{t''_2(s_2)t_4(s_2)}{t'_2(s_2)^3} +O(f^{-2}).
\end{equation}
We also have to expand $1/(s-s_A)$:
\begin{equation}
  \frac1{s-s_A}=\frac1{s-s_2}+\frac{s_4}{(s-s_2)^2}+O(f^{-4})
  =\frac1{s-s_2}-\frac{t_4(s_2)}{t'_2(s_2)(s-s_2)^2}+O(f^{-4}),
\end{equation}
and similarly for $1/(z_0-s_A)$, where we have
taken into account that $t'_2(s_2)s_4=-t_4(s_2)+\mo(f^{-6})$.
Hence, we find that
\begin{eqnarray}
  PC(1/t)=\frac1{t'(s_A)}
  \left(\frac1{s-s_A}-\frac1{s-s_2}\right)=
  \frac1{t'_2(s_2)}\left(\frac1{s-s_2}+\frac1{z_0-s_2}\right)
  -\frac{t_4(s_2)}{t'_2(s_2)^2}
  \left(\frac1{(s-s_2)^2}-\frac1{(z_0-s_2)^2}\right)
  \\
  -\left(
    \frac{t'_4(s_2)}{t'_2(s_2)^2}-
    \frac{t''_2(s_2)t_4(s_2)}{t'_2(s_2)^3}
  \right)
  \left(\frac1{s-s_2}-\frac1{z_0-s_2}\right)+O(f^{-2})
  =PC(1/t_2)-PC(t_4/t_2^2)+O(f^{-2}),
  \nonumber
\end{eqnarray}
so the total pole contribution
$PC(1/t)-PC(1/t_2)+PC(t_4/t_2^2)=O(f^{-2})$, which yields a
$O(f^{-6})$ contribution to $t(s)$.

\end{document}